\documentclass[%
 aip,
 pof,
 amsmath,amssymb,
reprint,
]{revtex4-1}
\usepackage{graphicx}
\usepackage{dcolumn}
\usepackage{bm}
\usepackage{color}
\usepackage{hyperref}
\usepackage{upgreek}

\usepackage{makecell}
\begin{document}

\title{Dynamics of cavitation bubbles inside a small corner}

\author{Jinzhao Liu}
\affiliation{State Key Laboratory of Engines, Tianjin University, Tianjin, 300350, China.}%
\author{Tianyou Wang}
\affiliation{State Key Laboratory of Engines, Tianjin University, Tianjin, 300350, China.}%
\affiliation{National Industry-Education Platform of Energy Storage, Tianjin University, Tianjin, 300350, China}
\author{Zhizhao Che}
\email{chezhizhao@tju.edu.cn}
 \affiliation{State Key Laboratory of Engines, Tianjin University, Tianjin, 300350, China.}
 \affiliation{National Industry-Education Platform of Energy Storage, Tianjin University, Tianjin, 300350, China}
\date{\today}

\begin{abstract}
Cavitation is a ubiquitous phenomenon in nature and bubble dynamics in open spaces have been widely studied, but the effects of the wall on the dynamics of cavitation bubbles in confined spaces are still unclear. Here, the dynamics of cavitation bubbles in small corners is studied experimentally, focusing on the interaction of the bubble with the wall. High-speed photography is used to visualize the temporal development of laser-induced cavitation bubbles in a small corner formed by two rigid walls, and the bubble spreading on the corner walls and the bubble migration is analyzed via digital image processing. We identify three distinct modes of bubble collapse, namely, collapse dominated by annular shrinkage, collapse dominated by wall attraction, and collapse governed by a combination of both annular shrinkage and wall attraction. The distribution of different collapse modes at different opening angles of the corner is also analyzed. The displacement and shrinkage of different parts of the bubble surface, as well as the direction and amount of bubble migration, are determined. The results show that the asymmetric structure of the corner leads to asymmetric bubble dynamics, including asymmetric bubble expansion, spreading, contraction, and migration.
\end{abstract}

\maketitle

\onecolumngrid
\section{Introduction}\label{sec:1}
Cavitation is the process of explosive formation and collapse of vaporous bubbles in a liquid caused by pressure reduction or energy deposit \citep{brennen13}. It is ubiquitous in nature and of great significance in numerous fields of science and engineering such as marine science, ocean engineering, mechanical and material engineering, environmental and chemical engineering, and medicine and life science. Therefore, the bubble dynamics in cavitation processes, as a fundamental problem of fluid mechanics, is attracting widespread interest. Cavitation also often occurs in narrow spaces. For example, cavitation configurations within parallel gaps include cavitation in underwater laser-assisted machining \citep{liao10}, electrical discharge machining \citep{shervanitabar06}, and biological applications \citep{hsiao13, mohammadzadeh17, yuan15}.

There have been many studies about the dynamics of cavitation bubbles and the induced jets near a single solid boundary \citep{blake82, blake87, chapman71, dorsaz16}. During this process, the collapse of cavitation bubbles forms high-speed jets directed towards the boundary, which is considered as one of the main reasons for the damage to the boundary. Cavitation shock wave is another cause of surface damage, as the collapse of bubbles can generate high-pressure waves in the order of MPa \citep{wang1994shock}. To investigate the mechanism of surface damage caused by cavitation shock waves, researchers have conducted numerical simulations \citep{gong2024numerical} and experimental studies \citep{philipp1998cavitation}, and found that the largest erosive force is caused by the collapse of a bubble in direct contact with the boundary, where pressures of up to several GPa act on the material surface. Philipp and Lauterborn \citep{lauterborn98} investigated the bubble-collapsing process on a flat metal specimen with high-speed photography. They found that when the bubble is generated at a distance less than twice of the maximum radius from a solid boundary, the jet velocity could rise to up to 83 m/s. Bubbles are accelerated towards the boundary by the Bjerknes forces during the collapse phases. Depending on the magnitude of pressure exerted on the wall, Luo \emph{et al.} \citep{luo18} categorized collapses of cavitation bubbles near the wall into three groups: primary impact area collapses, secondary impact area collapses, and slow release area collapses. For a bubble created very close to a rigid wall, different types of axial-jets are formed through annular-liquid-flow collision \citep{lechner20}. The surface forces due to the presence of these jets from cavitation bubbles have also been widely investigated experimentally \citep{dijkink08, occhicone19} and numerically \citep{dijkink18, li16}.

A bubble collapsing inside a narrow gap between two rigid walls experiences even richer dynamics than near a single solid boundary. For bubbles created at different locations of a narrow gap, three types of distinct jetting behaviors can occur: the transferred jet impacting on the distant wall, the double jet as a result of a bubble splitting and impacting on both walls, and the directed jet from a conically shaped bubble impacting on the closer wall \citep{gonzalezavila20}. The gap height is an important parameter for the bubble dynamics. With decreasing the gap height by up to 50\%, the bubble lifetime increases whereas the maximum projected bubble radius remains constant \citep{quinto-su09}. Under different gap heights, Gonzalez-Avila \emph{et al.} \citep{gonzalezavila11} observed three scenarios: neutral collapse at the gap center, collapse onto the lower wall, and collapse onto the upper wall. The bubble tends to collapse at the closer wall when the distance between the bubble and rigid walls is larger than the bubble diameter at its maximum expansion \citep{brujan19}. However, it was also shown that when the distance between the bubble and a wall is much smaller than the bubble's maximum radius, the bubble may collapse on the opposite wall \citep{gonzalezavila20Jetting}. A bubble can also be created between two parallel plates using a low-voltage electrical spark circuit \citep{quah18}. The radii and oscillation periods of spark-generated bubbles are one order of magnitude larger than those generated by laser beam \citep{cui15, khoo09}. Azam \emph{et al.} \citep{azam13} used two high-speed cameras to observe the behavior of a bubble from the side and the front simultaneously in a narrow gap. They found the presence of thin liquid films between each of the plates and the bubble throughout the bubble's lifetime. It results in bubble dynamics that are unaffected by the hydrophobic or hydrophilic nature of the plate surface.

Compared with bubbles in parallel gaps, much less research has been concerned with bubble dynamics in non-parallel gaps, a point of interest since the geometries in applications are often more complex than just a rigid boundary. For bubbles generated between a pair of symmetrically arranged oblique plates, Sagar and Moctar \citep{sagar23} observed bubbles with various shapes, including vertical pillar-shaped cavities and floating toroids. For a bubble in a right-angled corner, Brujan \emph{et al.} \citep{brujan18} studied the jetting behavior of the bubble, bubble migration, the formation and motion of the toroidal bubble, etc., and considered the influence of the stand-off distances between the bubble and the walls. White \emph{et al.} \citep{white23} quantified the pressure fields produced by a single bubble collapsing near two perpendicular rigid walls. In contrast to a bubble collapsing near a single wall, the collapse of bubbles within a critical stand-off distance is not symmetric about the bisecting plane due to the interaction between the bubble and the second wall. Wang \emph{et al.} \citep{wang20} investigated bubble dynamics at different angles of corners formed by two flat rigid boundaries. A jet forms toward the end of the collapse, pointing to the corner when a bubble initiates at the bisector of the corner but pointing to the near wall and inclined to the corner when a bubble initiates near one of the two walls. For a bubble at asymmetric positions in the corner, it firstly migrates to the near wall during the collapse since the Bjerknes attraction from the near wall is dominant \citep{cui20}. Recently, the bubbles near a corner formed by mixed boundaries have been the focus of research in many fields \citep{cui23}. For example, Li \emph{et al.} \citep{li19} studied the vertically neutral collapse of a bubble near a vertical rigid wall below a free surface.

Bubble generation inside small corners is a common phenomenon, such as cavitation occurring frequently in the corner formed by propeller blades. However, the study of bubble generation in corners has received relatively little attention, especially in small corners. Therefore, this paper conducts experimental research on this less attended area to effectively reveal the dynamics of bubbles generated within a small corner. This study reports experimental investigations on the bubble dynamics by focusing on the interaction of the bubble with the wall of the corner and indentifies three modes of bubble collapse. High-speed photography is used to visualize the temporal development of the bubble shape and behavior. The spreading and migration characteristics of the bubble during the bubble oscillation cycle is analyzed. The results show that the asymmetric structure of the small corner leads to asymmetric bubble dynamics.

\section{Experimental method}\label{sec:2}

\begin{figure}
  \centerline{\includegraphics[scale=0.6]{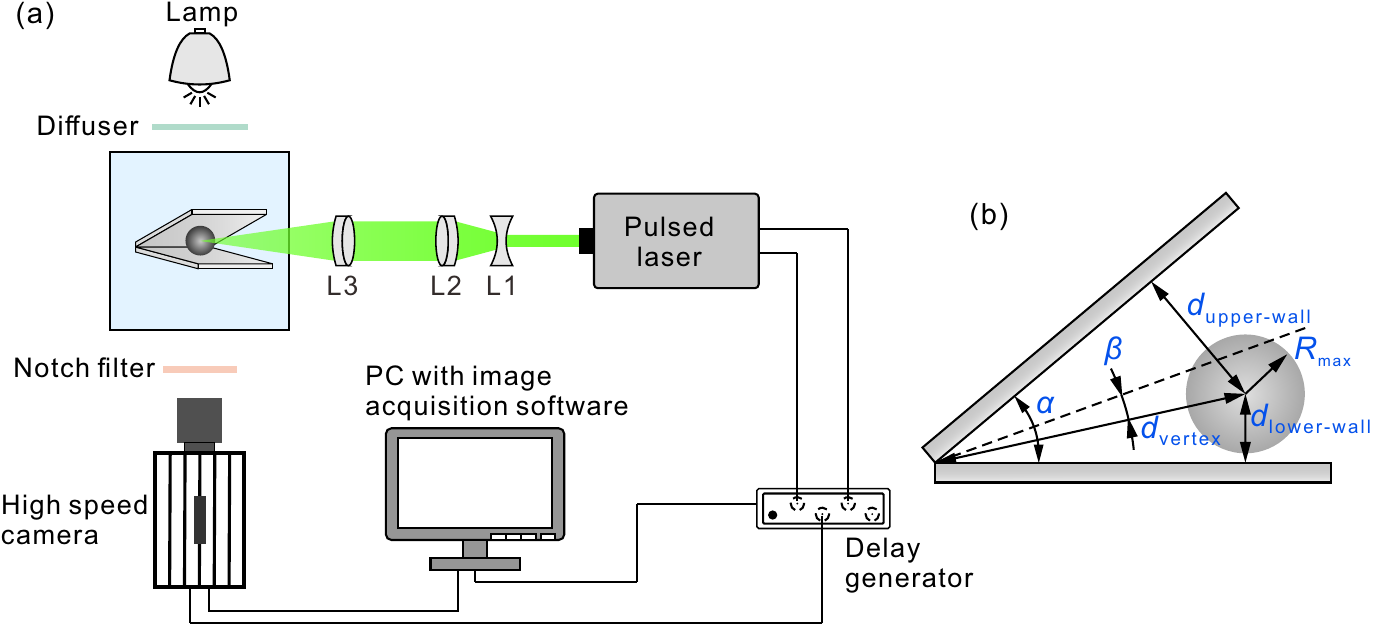}}
  \caption{(a) Schematic diagram of the experimental setup. (b) Schematic drawing of the parameters for the initial bubble position in the corner.}
\label{fig:01}
\end{figure}
The experimental setup is schematically shown in Figure \ref{fig:01}a. The cavitation bubble was generated by a Q-switched Nd:YAG laser pulse (Nano L 135-15 PIV, Litron). The laser was collimated with a combination of a plano-concave lens (L1) and a plano-convex lens (L2). After collimation, the laser beam was focused by a plano-convex lens (L3) into an acrylic cuvette filled with distilled water. The cuvette is made of 5-mm thick acrylic glass plates, and has height $H_1 = 150$ mm, length $H_2 = 150$ mm, and width $W = 150$ mm. The acrylic plates behave as rigid walls without any noticeable deformation in the experiment. The focused laser pulse ionizes the water, generating a breakdown plasma \citep{vogel96}. This facilitated obtaining a single uninterrupted plasma, thereby avoiding the generation of multiple bubbles \citep{han15}. The wavelength of the laser light is 532 nm and the maximum energies of the laser pulses are up to 135 mJ with a duration of 6 ns. Images of bubble evolution were recorded by a high-speed camera (Phantom TMX 7510) at a frame rate of 375,000 frames per second (fps) with a resolution of $640 \times 320$ pixels and an exposure time of 2.3 $\upmu$s. The interval between two image frames is 2.7 $\upmu$s, which is much smaller than the period of bubble oscillation (about 200 $\upmu$s). A notch filter was placed in front of the camera to prevent scattered laser beams from damaging the camera sensor. An LED lamp was used to provide mildly diffused light with a diffuser. A digital delay generator (Quantum Composer 9214) was used to trigger the pulsed laser and the high-speed camera synchronously.

The corner used in the experiment was formed by two transparent rectangular acrylic glass plates with a thickness of 5 mm. The two acrylic glass plates were placed at an angle of $\alpha = 10 ^\circ$. The angle formed by the bisector and the line of the initial bubble center and the corner vertex is denoted as $\beta$. The energy of the pulsed laser was set constant at about 12.4 mJ, and the maximum radius of the laser-induced bubble in the experiment was $R_{\max} = 0.70 \pm 0.05$ mm. Since the bubble at its maximum expansion is non-spherical, the equivalent spherical radius $R_{\max}$ is obtained via image processing, which represents the radius of a spherical bubble with the same volume. It was calculated using the area of the bubble $A_{\max}$ in the image at its maximum expansion, i.e., ${{R}_{\max }}=\sqrt{{{{A}_{\max }}}/{\pi }}$.

The relative positions of the bubble with respect to the rigid walls of the corner are important parameters for the bubble evolution process. For a bubble near a single rigid boundary in previous studies, its dynamics can be characterized by the stand-off distance \citep{blake87, lauterborn75, plesset71}. Following a similar concept, for a bubble near a corner in this study, we use three non-dimensional parameters to depict the initial position of the bubble in the corner, namely the dimensionless upper-wall distance $d_{\text{upper-wall}}^{ {*}}={{{d}_{\text{upper-wall}}}}/{{{R}_{\max }}}$, dimensionless lower-wall distance $d_{\text{lower-wall}}^{ {*}}={{{d}_{\text{lower-wall}}}}/{{{R}_{ {\max}}}}$, and the dimensionless vertex distance $d_{\text{vertex}}^{{*}}={{{d}_{\text{vertex}}}}/{{{R}_{{\max}}}}$, where ${{d}_{\text{upper-wall}}}$ and ${{d}_{\text{lower-wall}}}$ are, respectively, the distance from the bubble initial position to the upper wall and the lower wall, and ${{d}_{\text{vertex}}}$ is the distance between the bubble initial position and the vertex of the corner, as illustrated in Figure \ref{fig:01}b.

\section{Results and discussion}\label{sec:3}
\subsection{Behaviors of cavitation bubbles at a corner}\label{sec:3.1}
According to the characteristics of the bubble evolution and their underlying mechanisms, we classify the modes of bubble collapse into three main categories: collapse dominated by annular shrinkage, collapse dominated by wall attraction, and collapse governed by a combination of both annular shrinkage and wall attraction. We next discuss these three categories.

\subsubsection{Annular shrinkage}\label{sec:3.1.1}

\begin{figure}
  \centerline{\includegraphics[scale=0.9]{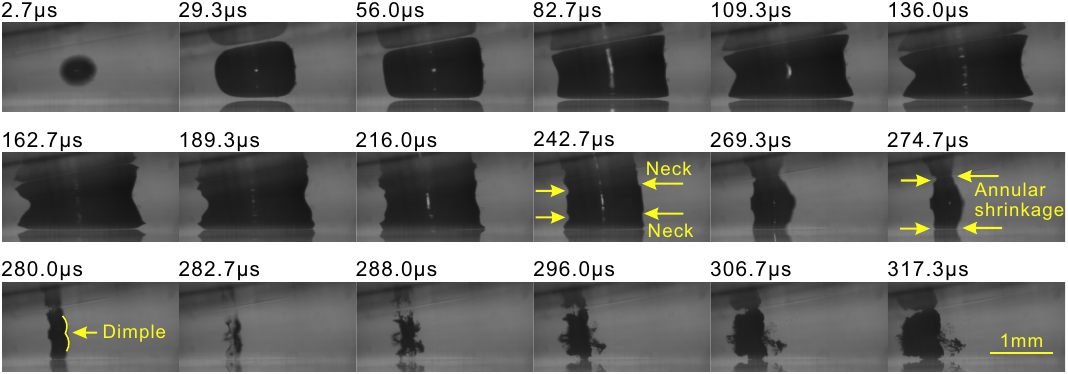}}
  \caption{Image sequence of bubble collapse dominated by annular shrinkage (Multimedia available online). Here, $d_{\text{vertex}}^{*} = 6.87$, $d_{\text{lower-wall}}^{ {*}} = 0.65$, and $R_{\text{max}} = 0.72$ mm at $t = 136.0 \upmu$s. }
\label{fig:02}
\end{figure}

Figure \ref{fig:02} (Multimedia available online) shows a scenario where the bubble collapse is dominated by annular shrinkage. During the growth phase, due to the constraining effect of the walls, the top and bottom of the bubble are flattened. However, the expansion in the middle of the bubble is relatively less pronounced compared to the top and bottom sections. During the collapse phase, two concave annular rings form on the upper and lower parts of the bubble. These rings develop into `necks' on the bubble during subsequent stages of the collapse process (242.7 $\upmu$s). Meanwhile, the top and bottom of the bubble shrink faster than the middle. This results in an oval shape of the bubble with smaller top and bottom regions and a larger central region during the late stages of the collapse (274.7--282.7 $\upmu$s). Then, the bubble detaches from the wall at the top and bottom, undergoes fragmentation, and enters the rebound phase. Since the bubble shrinks mainly in the annular direction in the collapse process, we refer to this phenomenon as bubble collapse dominated by annular shrinkage. Even though there is interaction of the bubble with the wall, the interaction is not very strong and the bubble does not make complete contact with the wall, and there exists a liquid film between the bubble surface and the wall. Such liquid film has been confirmed to exist during the bubble evolution near a single wall, and its thickness is on the micrometer scale \citep{reuter19}. The liquid within the film flows outward, generating a strong shear stress and creating a substantial pressure gradient in the region where the bubble approaches the wall. This results in intense annular shrinkage on the bubble's top and bottom. Similarly, the phenomenon of intense constriction also occurs when bubbles are formed very close to a rigid wall or within the gap between two parallel plates \citep{lechner20, zeng20}. In the final stage before the bubble's rupture (280.0 $\upmu$s), a slight dimple appears on the right-middle of the bubble. Subsequently, the bubble breaks apart during the rebound phase, producing numerous tiny bubbles (see the dark region in the image). And these tiny bubbles migrate to the left (see the dark region expanding to the left in the images in 288.0--317.3 $\upmu$s). The dimple on the right side before the bubble breaks and the movement of the bubble toward the left after the bubble breaks suggest that, during the collapse, the bubble is likely influenced by the jet pointing towards the corner vertex. The generation of the jet is due to the higher pressure of the fluid on the right side of the bubble compared to the left and this asymmetry in pressure is a consequence of the asymmetric corner structure. Simultaneously, unlike the top and bottom regions, the middle of the bubble does not experience significant annular shrinkage. As a result, the influence of the jet directed towards the corner vertex is more pronounced in the middle of the bubble. This leads to a dimple on the right-middle of the bubble, causing it to collide with the opposing left surface.

\subsubsection{Wall attraction}\label{sec:3.1.2}

\begin{figure}
  \centerline{\includegraphics[scale=0.8]{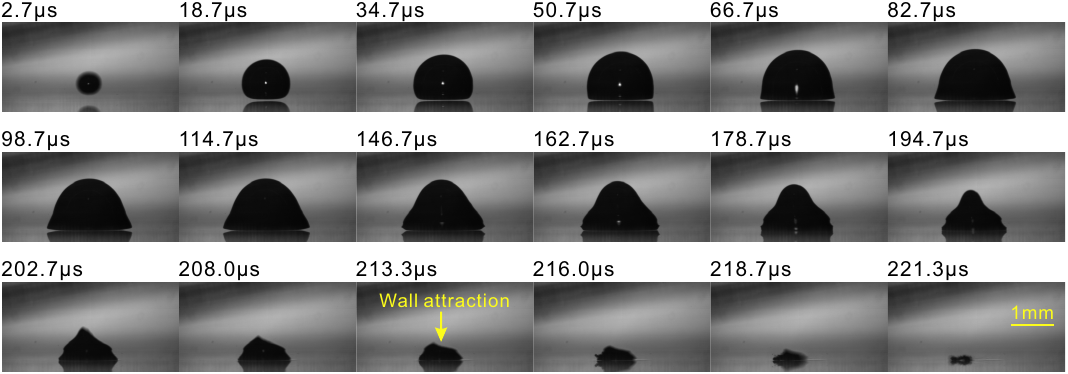}}
  \caption{Image sequence of bubble collapse dominated by wall attraction (Multimedia available online). Here, $d_{\text{vertex}}^{*}$ = 11.61,  $d_{\text{lower-wall}}^{\text{*}}$ = 0.54, and $R_{\text{max}} = 0.72$ mm at $t = 98.7 \upmu$s.}
\label{fig:03}
\end{figure}

When the bubble is distant from the corner vertex and close to one of the side walls, the phenomenon of wall attraction occurs, as shown in Figure \ref{fig:03} (Multimedia available online). In this case, the evolution of the bubble is similar to that of a bubble near a single rigid wall \citep{vogel89}, and a ring-like `neck' appears in the lower portion of the bubble during its collapse. In contrast, since the bubble is farther away from the upper wall, the attractive influence of the upper wall on the bubble is much weaker compared to that of the lower wall. Even with the presence of a ring-like constriction at the bottom of the bubble, the attractive force exerted by the lower wall remains dominant. This results in the bubble being unable to detach from the lower wall, ultimately leading to its collapse directly against the lower wall (213.3--221.3 $\upmu$s). In addition, due to the influence of the asymmetry in the corner structure, the bubble shape during the bubble collapse process is not axisymmetric. This can be evidenced by the gradual leftward movement of the protrusion at the top of the bubble (94.7--208.0 $\upmu$s).

\subsubsection{Combined annular shrinkage and wall attraction}\label{sec:3.1.3}

\begin{figure}
  \centerline{\includegraphics[scale=0.8]{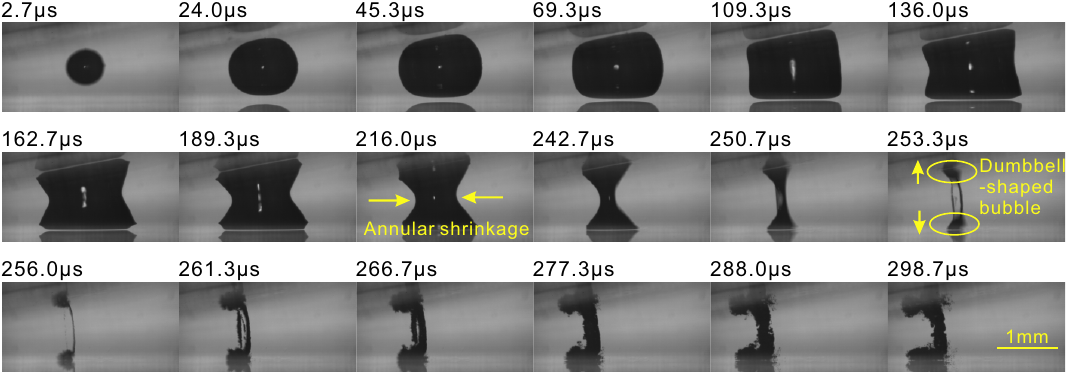}}
  \caption{Image sequence of bubble collapse dominated by both annular shrinkage and wall attraction (Multimedia available online). Here, $d_{\text{vertex}}^{*}$ = 8.47, $d_{\text{lower-wall}}^{{*}}$ = 0.75, and $R_{\text{max}} = 0.71$ mm at $t = 136.0 \upmu$s. }
\label{fig:04}
\end{figure}

When the bubble is generated close to the bisector of the corner and the distances from both the upper and lower wall surfaces are slightly greater compared to the case with annular shrinkage, the process of bubble collapse is governed by a combination of both annular shrinkage and wall attraction, as shown in Figure \ref{fig:04} (Multimedia available online) for $d_{\text{vertex}}^{*}$ = 8.47 and $d_{\text{lower-wall}}^{ {*}}$ = 0.75. In the initial stages of bubble formation, the bubble exhibits a regular spherical shape (2.7--24.0 $\upmu$s). As the bubble expands and encounters constraints from the upper and lower walls, the top and bottom of the bubble become flattened (69.3--109.3 $\upmu$s). During the contraction process, there is no noticeable annular shrinkage along the wall surfaces for either the top or bottom of the bubble. Instead, the annular shrinkage is more prominent in the middle of the bubble (136.0--242.7 $\upmu$s). Then, the bubble is attracted to the upper and lower walls, leading to a rupture at the thin middle part and thus initiating the collapse. In this condition, the primary effect on the top and bottom of the bubble is the attraction exerted by the upper and lower walls. This results in the bubble transforming into a dumbbell shape (253.3 $\upmu$s). The gas column in the middle of the bubble bends to the side, resembling a bow shape. The bubble deforms into a bow shape because the middle of the bubble contracts faster than the top and bottom, and the right side of the top and bottom of the bubble contracts faster than that on the left side. That is, when the middle of the bubble contracts to its smallest and becomes a column, the top and bottom are still in the process of contracting. Due to the influence of an asymmetric corner structure, there is a greater degree of contraction on the right sides of the top and bottom of the bubble, causing the top and bottom to be positioned more to the left relative to the middle, thus the gas column in the middle of the bubble assumes a bow shape. The top and bottom of the bubble remain in contact with the walls throughout the entire collapse cycle. After the bubble expands to its maximum volume, the internal pressure of the bubble is lower than the surrounding fluid pressure, leading the bubble into a contraction phase. When the bubble contracts to its smallest size, the bubble interface becomes unstable, causing the bubble to burst into many smaller bubbles. Similarly, due to the asymmetric structure created by the corner, the fluid pressure on the right side of the bubble is higher than that on the left. Thus, during the rebound phase, many small fragmented bubbles gradually migrate towards the left (261.3--298.7 $\upmu$s).

\subsection{Effects of initial bubble position}\label{sec:3.2}
\subsubsection{Overview of the bubble behavior within the corner for $\alpha = 10 ^\circ$}\label{sec:3.2.1}

\begin{figure}
  \centerline{\includegraphics[scale=0.6]{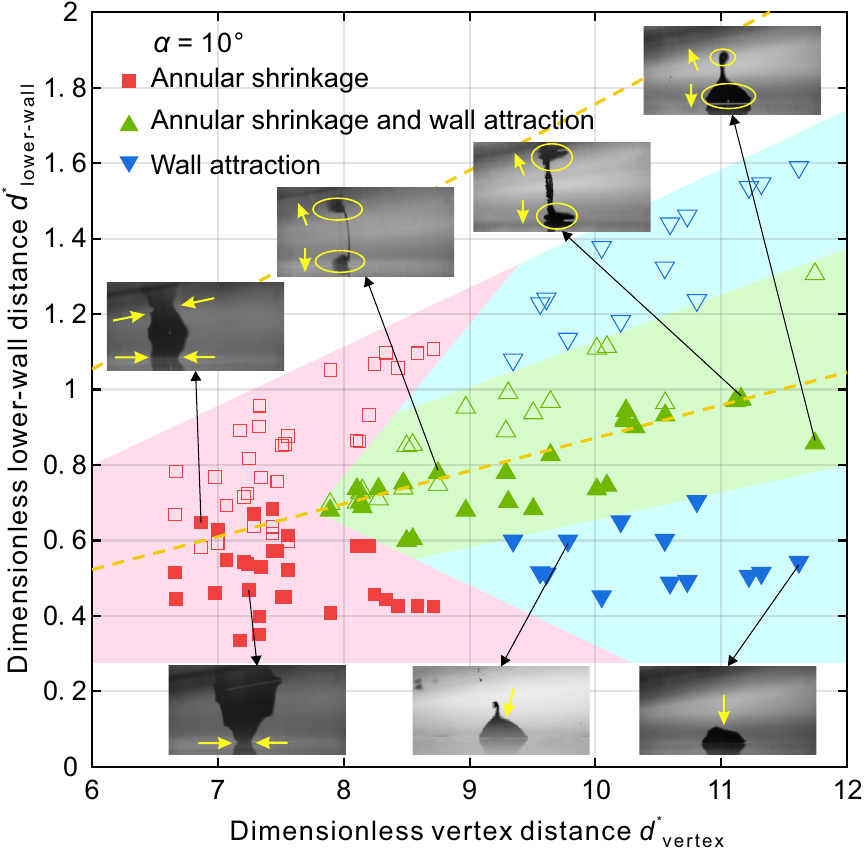}}
  \caption{Distribution of different modes of bubble collapse within a small corner of $\alpha = 10 ^\circ$. Solid symbols represent data obtained from the experiment, whereas hollow symbols represent data obtained by symmetry reflection of the experimental results about the corner bisector. The yellow arrows indicate the contraction directions of the bubble surfaces. The yellow dashed line indicates the positions of the upper wall and the corner bisector.}
\label{fig:05}
\end{figure}

The results in Section \ref{sec:3.1} have shown that the initial bubble position is important for the evolution of cavitation bubbles. To study the effects of the initial bubble position, we perform experiments by generating bubbles at different positions of the corner, as shown in Figure~\ref{fig:05}. The initial bubble position can be characterized by the dimensionless lower-wall distance $d_{\text{lower-wall}}^{ {*}}$ and dimensionless vertex distance $d_{\text{vertex}}^{*}$. Since the corner structure is symmetric with respect to the bisector line, for two bubbles within the corner that are symmetrically positioned with respect to the bisector, their behavior should also be symmetric with each other. The effect of gravity can be quantified using the a dimensionless parameter proposed by Blake \citep{blake1988kelvin}, $\delta = \sqrt {{{\rho g{R_{\max }}} \mathord{\left/ {\vphantom {{\rho g{R_{\max }}} {({p_0} - {p_v})}}} \right. \kern-\nulldelimiterspace} {({p_0} - {p_v})}}} $, where $\rho $ is the liquid density, $g$ is the gravitational acceleration, ${R_{\max }}$ is the maximum radius of the bubble, ${p_0}$ is the pressure at infinity at the vertical position of the bubble center and ${p_v}$ is the vapor pressure. In our experiment, the dimensionless parameter $\delta $ is very small, approximately 0.0083, indicating that the influence of gravity can be neglected. There have been some studies about the effect of gravity on the cavitation dynamics. It was found that the effects of gravity are important for the generation of micro-jets \cite{Obreschkow2011gravity}, which occurs at the end of the collapse stage. Since we focus on interaction of the bubble with the walls during the expansion and contraction of the bubble, the effect of the gravity can be neglected. Therefore, we conducted experiments to generate bubbles only within the region between the bisector and the lower wall. The behavior of bubbles in the region between the bisector and the upper wall can be deduced based on symmetry considerations. In Figure \ref{fig:05}, solid symbols represent data obtained from the experiment directly, whereas hollow symbols represent data obtained by symmetry reflection of the experimental results about the corner bisector. Additionally, because bubbles cannot be generated very close to the wall in the experiment, there is no data point for very small $d_{\text{lower-wall}}^{ {*}}$ or its reflection.

The bubble collapse behavior in a small corner of $\alpha = 10 ^\circ$ is summarized in Figure \ref{fig:05}. For small values of $d_{\text{vertex}}^{*}$ (see the region with red squares in Figure \ref{fig:05}), the bubble is very close to both the upper and lower walls. Even when the bubble is generated at the bisector, there is strong annular shrinkage at the top and bottom of the bubble, leading to the bubble detaching from the walls. When the bubble approaches one wall of the corner, the annular shrinkage becomes even more pronounced in the part of the bubble that is in contact with the wall. Consequently, the bubble collapse is dominated by annular shrinkage. As $d_{\text{vertex}}^{*}$ increases and when the bubble is generated near the bisector line (see the region with green upper triangles in Figure \ref{fig:05}), annular shrinkage occurs in the middle of the bubble, while the bubble experiences an attractive force from both the upper and lower walls. As a result, the bubble transforms into a dumbbell shape. For larger values of $d_{\text{vertex}}^{*}$ and when the bubble is generated close to one wall (see the region with blue lower triangles in Figure \ref{fig:05}), the bubble is primarily influenced by the attraction from the wall. This causes the bubble to completely collapse towards the wall.

\subsubsection{Effect of dimensionless lower-wall distance $d_{\mathrm{lower-wall}}^{{*}}$}\label{sec:3.2.2}

\begin{figure}
  \centerline{\includegraphics[scale=0.8]{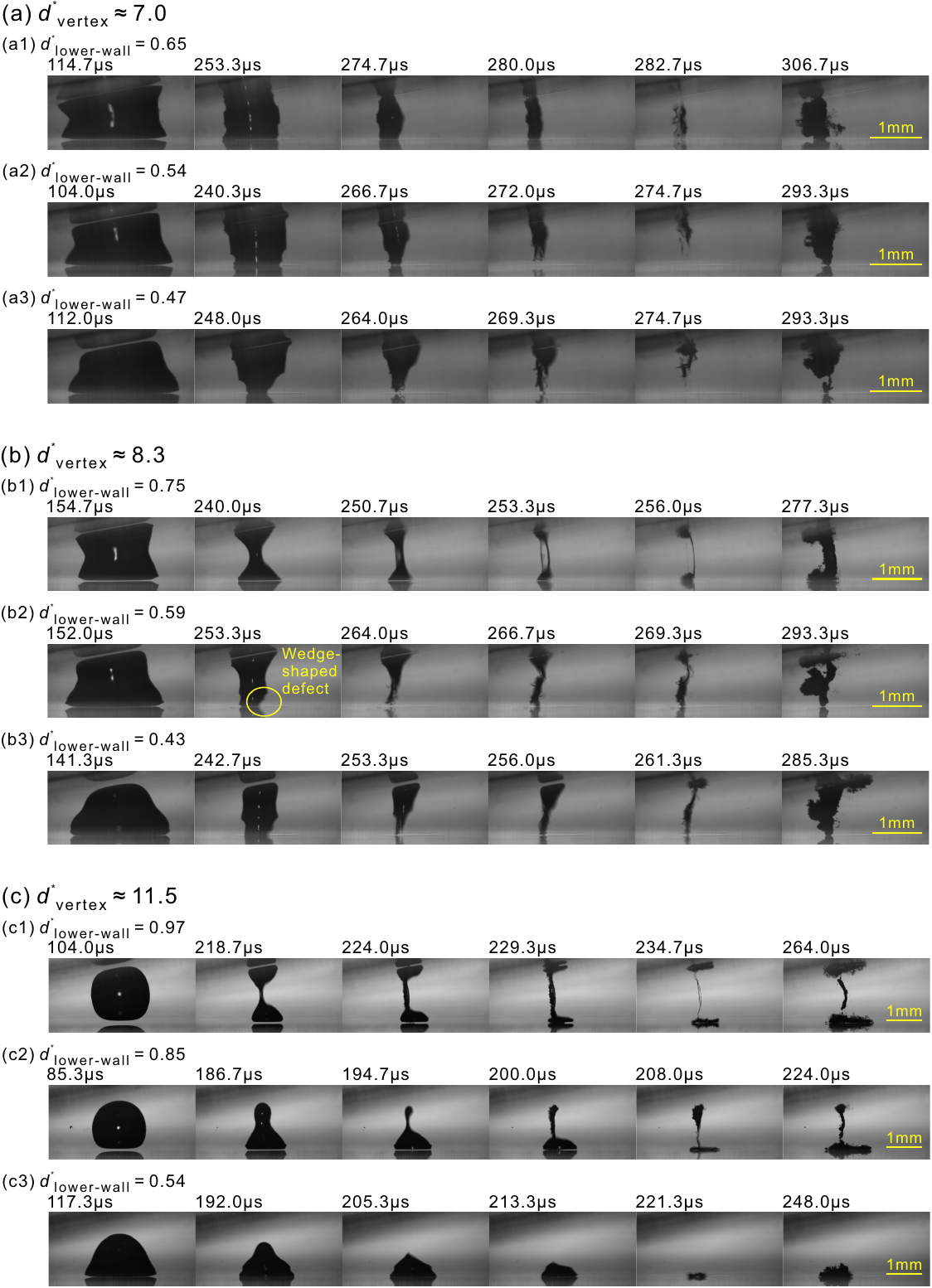}}
  \caption{Bubble behaviors as the initial position gradually approaches the lower wall, under three different values of $d_{\text{vertex}}^{*}$. (a) $d_{\text{vertex}}^{*}\approx 7.0$ (Multimedia available online); (b) $d_{\text{vertex}}^{*}\approx 8.3$ (Multimedia available online); (c) $d_{\text{vertex}}^{*}\approx 11.5$ (Multimedia available online).}
\label{fig:06}
\end{figure}

To further explain the effect of the lower-wall distance, image sequences are presented in Figure \ref{fig:06} to show the variation of the bubble behavior as the initial bubble position gradually approaches the lower wall, under three different values of $d_{\text{vertex}}^{*}$.

When $d_{\text{vertex}}^{*}$ is relatively small (e.g., $d_{\text{vertex}}^{*} \approx 7.0$ in Figure \ref{fig:06}a, Multimedia available online), as $d_{\text{lower-wall}}^{ {*}}$ decreases, the annular shrinkage at the bottom of the bubble becomes more intense, whereas the annular shrinkage at the top of the bubble gradually diminishes. The extent of contraction at the bottom of the bubble is significantly greater than that at the top and middle, resulting in the bubble assuming an inverted triangle shape during the late stages of bubble collapse. This results in a gradual upward migration of the bubble's centroid. As the bubble continues to contract, the bottom of the bubble will detach from the lower wall earlier than the top wall. Then tiny fragmented microbubbles appear at the bottom of the bubble, while the rest of the surface contours of the bubble remains smooth. Consequently, fragmentation of the bubble commences at the bottom and proceeds upwards gradually. Ultimately, the bubble shrinks to its smallest volume and collapses against the upper wall. During the rebound phase, under the influence of the jet directed towards the vertex, the bubble cluster continues to migrate leftward.

The image sequences in Figure \ref{fig:06}b (Multimedia available online) show the collapse of bubbles for a moderate dimensionless vertex distance (i.e., $d_{\text{vertex}}^{*} \approx 8.3$). At a relatively large dimensionless lower-wall distance (i.e., $d_{\text{lower-wall}}^{ {*}} = 0.75$), a strong annular contraction occurs in the middle of the bubble, leading to the fragmentation of the bubble from its middle towards both the upper and lower ends. At a moderate dimensionless lower-wall distance (i.e., $d_{\text{lower-wall}}^{ {*}} = 0.59$), the annular shrinkage at the bottom of the bubble is more pronounced compared to the middle section, whereas there is virtually no discernible annular shrinkage at the top of the bubble. During the collapse phase, the bubble gradually migrates upward. At the end of the collapse stage, a wedge-shaped defect forms at the lower right corner of the bubble. This could be attributed to the influence of the asymmetric structure, which causes the annular shrinkage at the bottom of the bubble to proceed more rapidly on the right side than on the left side. The bottom of the bubble detaches from the lower wall first and begins to fragment upwards from the bottom, being consistent with the phenomena described for bubbles close to the vertex (e.g., $d_{\text{vertex}}^{*} \approx 7.0$ in Figure \ref{fig:06}a). For a relatively small dimensionless lower-wall distance (i.e., $d_{\text{lower-wall}}^{ {*}} = 0.43$), it is only after the fragmentation of the bubble that the fine and dispersed bubbles make contact with the upper wall. The annular shrinkage at the bottom of the bubble is more pronounced than at the top and the middle of the bubble. Similar to the case with $d_{\text{lower-wall}}^{ {*}} = 0.59$ in Figure \ref{fig:06}b, a wedge-shaped defect is present at the lower right corner of the bubble. Ultimately, the bottom of the bubble detaches from the lower wall and undergoes fragmentation from the bottom.

The bubble behavior at large dimensionless vertex distances (i.e., $d_{\text{vertex}}^{*} \approx 11.5$) is shown in Figure \ref{fig:06}c (Multimedia available online). At a relatively large dimensionless lower-wall distance (i.e., $d_{\text{lower-wall}}^{ {*}} = 0.97$), intense annular shrinkage occurs only in the central region of the bubble during the contraction of the bubble. Under the combined influence of the annular shrinkage in the middle and the attractive forces exerted by the upper and lower walls, the bubble transforms into a dumbbell shape. The middle of the bubble is a slender column and is offset to the left. This is because the asymmetric structure leads to unequal fluid pressures on the left and right sides of the bubble, consequently resulting in the asymmetric contraction of the bubble. This phenomenon is analogous to the bubble behavior in rectangular channels \citep{brujan22}, in which the structure connecting the upper and lower parts of the bubble has a slight offset due to the effect of the left wall. During the first growth-collapse cycle of the bubble, the top and bottom of the bubble remain consistently uncontacted with the walls. The fragmentation of the bubble initiates from the cylindrical portion. Subsequently, the surface of the bubble becomes unsmooth, giving rise to fine and dispersed bubbles. The fragmentation then progresses from the central region to both the upper and lower parts of the bubble. At a smaller dimensionless lower-wall distance (i.e., $d_{\text{lower-wall}}^{ {*}} = 0.85$), there is a distinct annular shrinkage noticeable in the middle of the bubble. However, because the distance from the bubble to the upper wall is greater than that to the lower wall, the attractive effect of the lower wall on the bubble is stronger at this moment. Consequently, the annular shrinkage in the middle of the bubble causes the bubble to transform into a conical shape. For a small dimensionless lower-wall distance (i.e., $d_{\text{lower-wall}}^{ {*}} = 0.54$), the attraction from the lower wall dominates the collapse of the bubble. This results in the bubble being unable to detach from the lower wall, ultimately leading to its collapse directly against that wall.

\subsubsection{Effect of dimensionless vertex distance $d_{\mathrm{vertex}}^{*}$}\label{sec:3.2.3}

\begin{figure}
  \centerline{\includegraphics[scale=0.8]{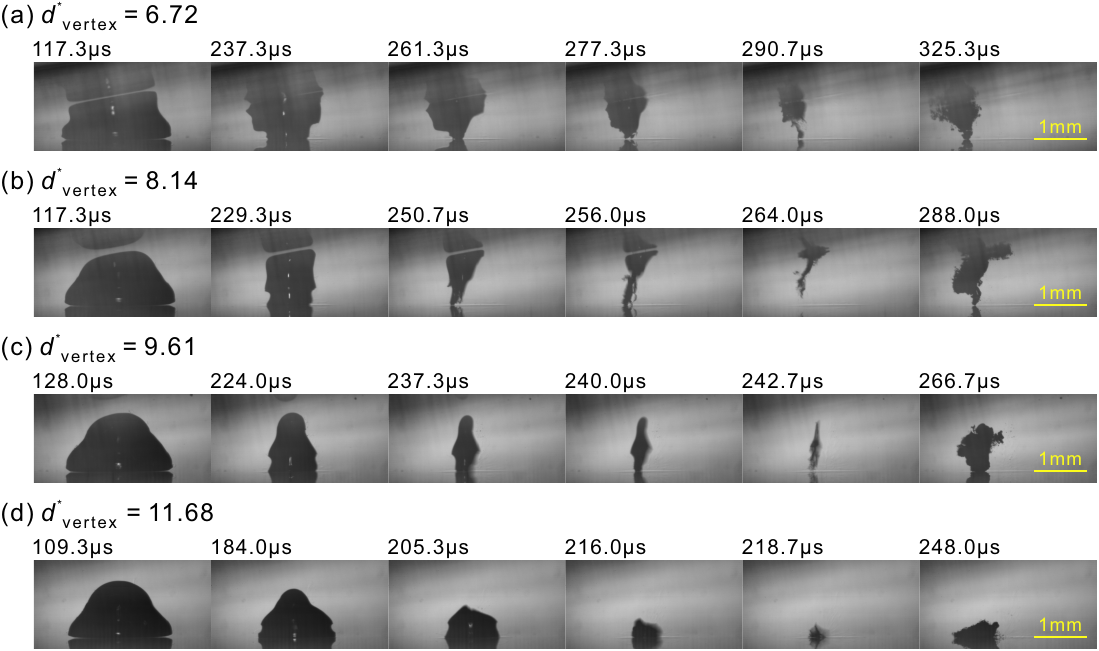}}
  \caption{Bubble behaviors as the initial position gradually moves away from the vertex of the corner (Multimedia available online). Here, $d_{\text{lower-wall}}^{*} \approx 0.44$. (a) $d_{\text{vertex}}^{*}=6.72$, (b) $d_{\text{vertex}}^{*}=8.14$, (c) $d_{\text{vertex}}^{*}=9.61$, and (d) $d_{\text{vertex}}^{*}=11.68$.}
\label{fig:07}
\end{figure}
To further investigate the effects of $d_{\text{vertex}}^{*}$ on bubble behavior, in our experiment, we maintained a constant distance between the initial position of the bubble and the lower wall (i.e., $d_{\text{lower-wall}}^{ {*}} \approx 0.44$) while progressively increasing the distance of the bubble from the corner vertex, as shown in Figure \ref{fig:07} (Multimedia available online). Through comparison, we can see that, as $d_{\text{vertex}}^{*}$ increases, the bubble becomes more difficult to detach from the lower wall. The annular shrinkage at the bottom of the bubble also gradually becomes less obvious.

For the collapse of bubbles in a corner, the strength of the annular shrinkage at the bottom of the bubble is determined by the distances of the bubble's initial position to the upper wall and to the lower wall. Among these two factors, the distance to the lower wall plays a dominant role. The closer the bubble forms to the lower wall, the more pronounced the annular shrinkage at the bubble's base becomes. Simultaneously, the upper wall exerts a certain degree of promoting effect on the annular shrinkage at the bottom of the bubble. The smaller the distance from the bubble to the upper wall (but not less than the distance between the bubble and the lower wall), the promoting effect becomes more pronounced. As shown in Figure \ref{fig:07}, while the distance to the lower wall remains constant, the distance to the upper wall increases as the bubble moves farther away from the vertex of the corner. This weakens the promoting effect of the upper wall on the annular shrinkage at the bottom of the bubble, causing the annular shrinkage to become increasingly indistinct and the bubble to become increasingly difficult to detach from the lower wall.

\subsection{Effect of the opening angle $\alpha $}\label{sec:3.3}

\begin{figure}
  \centerline{\includegraphics[scale=0.75]{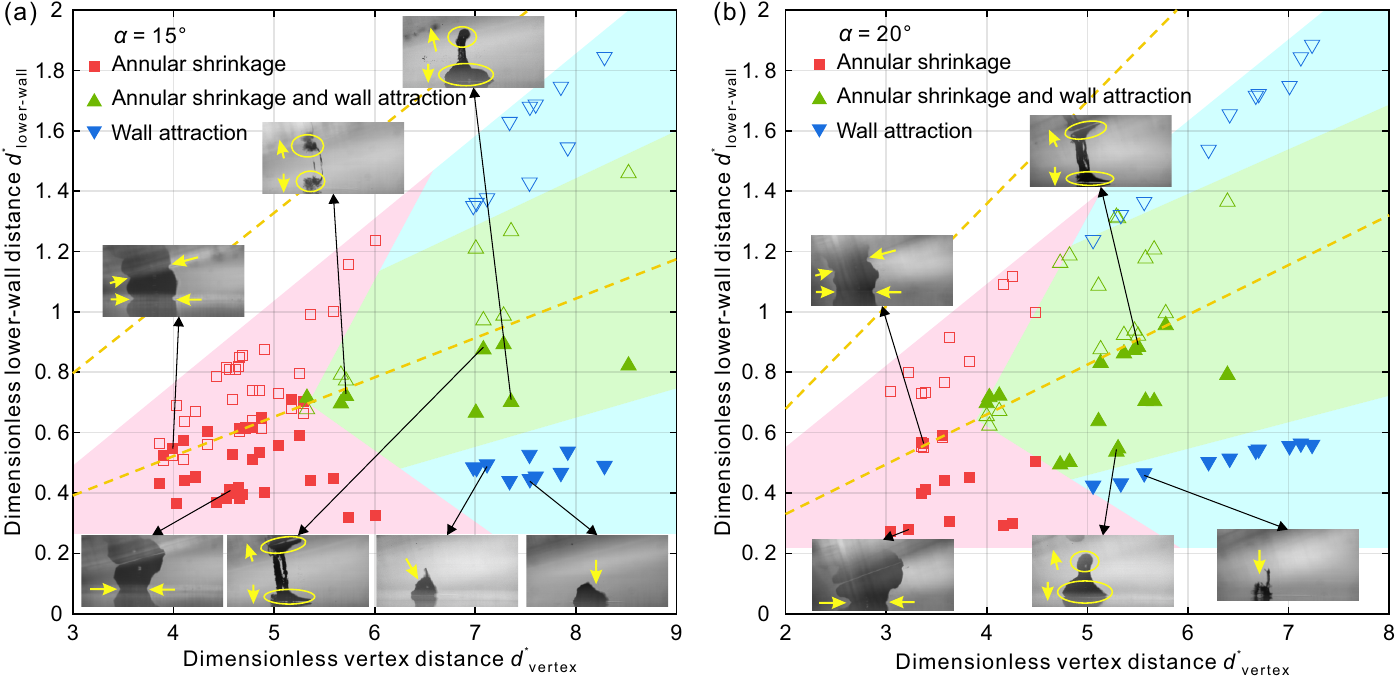}}
  \caption{Distributions of different modes of bubble collapse within small corners: (a) $\alpha = 15^\circ$, (b) $\alpha = 20^\circ$. The yellow arrows indicate the contraction directions of the bubble surfaces.}
\label{fig:08}
\end{figure}

To study the effect of the opening angle of the corner, we performed experiments with different opening angles at $\alpha = 15^\circ$ and $\alpha = 20^\circ$, and the results of bubble collapse behavior are shown in Figure \ref{fig:08}. Together with the results of the opening angle $\alpha = 10^\circ$ in Figure \ref{fig:05}, we can see that, as $\alpha $ increases, bubbles collapse dominated by annular shrinkage becomes more concentrated near the vertex of the corner. This trend is because, when the corner is larger, bubbles must be closer to the vertex to ensure that the distance from the bubble to both the upper and lower walls is small enough for the bubble to be strongly affected by both upper and lower walls simultaneously, which causes a noticeable annular shrinkage at the top or bottom of the bubble. Similarly, in a larger corner, when the initial position of the bubble is slightly away from the vertex, the distance from the bubble to both the upper and lower walls becomes greater than that in a smaller corner. Consequently, the bubble can escape the strong influence of the walls at a position closer to the vertex than in a smaller corner. Thus, the shifts of the bubble collapse mode from dominated by annular shrinkage to dominated by wall attraction, and also to dominated by combined annular shrinkage and wall attraction, occur at positions closer to the vertex than in smaller corners.

\begin{figure}
  \centerline{\includegraphics[scale=0.8]{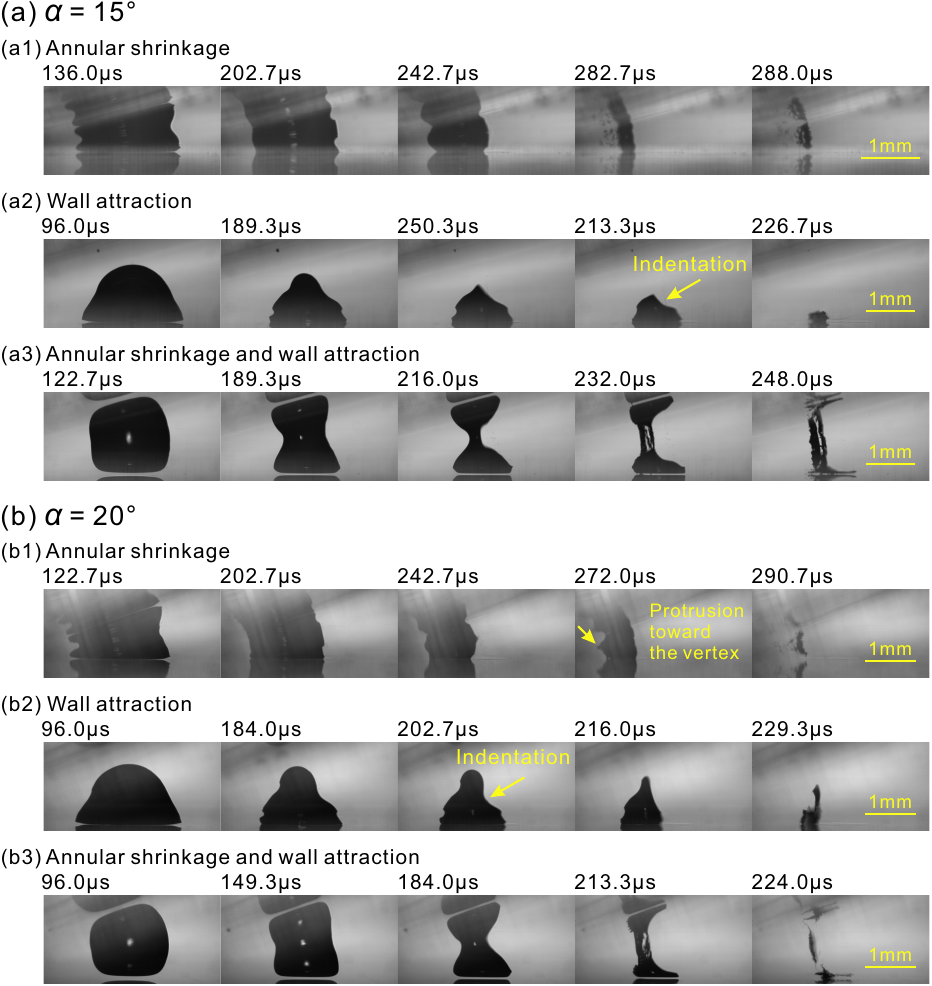}}
  \caption{Image sequence of bubble collapse for different collapse modes: (a) $\alpha = 15^\circ$ (Multimedia available online), (b) $\alpha = 20^\circ$ (Multimedia available online).
Here, (a1) $d^*_\text{vertex}=4.0$, $d^*_\text{lower-wall}=0.55$,
(a2) $d^*_\text{vertex}=7.5$, $d^*_\text{lower-wall}=0.44$,
(a3) $d^*_\text{vertex}=7.1$, $d^*_\text{lower-wall}=0.88$,
(b1) $d^*_\text{vertex}=3.4$, $d^*_\text{lower-wall}=0.56$,
(b2) $d^*_\text{vertex}=5.6$, $d^*_\text{lower-wall}=0.47$,
(b3) $d^*_\text{vertex}=5.5$, $d^*_\text{lower-wall}=0.88$.
}
\label{fig:09}
\end{figure}

Figures \ref{fig:09}a (Multimedia available online) and \ref{fig:09}b (Multimedia available online) show the collapse processes of bubbles within corners for $\alpha = 15^\circ$ and $\alpha = 20^\circ$, respectively. Compared with that for $\alpha = 10^\circ$ in Figure \ref{fig:05}, we can see that in larger corners, the collapse dominated by annular shrinkage occurs closer to the vertex. As a result, the attraction of the vertex to the bubble is more pronounced, leading to a protrusion toward the vertex of the corner in the later stage (272.0 $\upmu$s, Figure \ref{fig:09}b, for annular shrinkage mode). For the collapse mode dominated by wall attraction, the bubble in larger corners still contracts toward the lower wall and finally collapses on the lower wall. However, unlike in a corner of $\alpha = 10^\circ$, for corners of $\alpha = 15^\circ$ and $\alpha = 20^\circ$, the collapse dominated by annular shrinkage shifts to the collapse dominated by wall attraction at positions closer to the vertex. Consequently, during the collapse process, the bubble experiences a stronger attraction toward the vertex. This results in a greater extent of indentation on the right side of the bubble compared to the left side (213.3 $\upmu$s in Figure \ref{fig:09}a and 202.7 $\upmu$s in Figure \ref{fig:09}b, for wall attraction mode). For the collapse dominated by combined annular shrinkage and wall attraction, there is still an evident annular shrinkage in the middle of the bubble, and the bubble transforms into a dumbbell shape under the attraction of the upper and lower walls.

\subsection{Bubble spreading on the corner walls}\label{sec:3.4}

\begin{figure}
  \centerline{\includegraphics[scale=0.7]{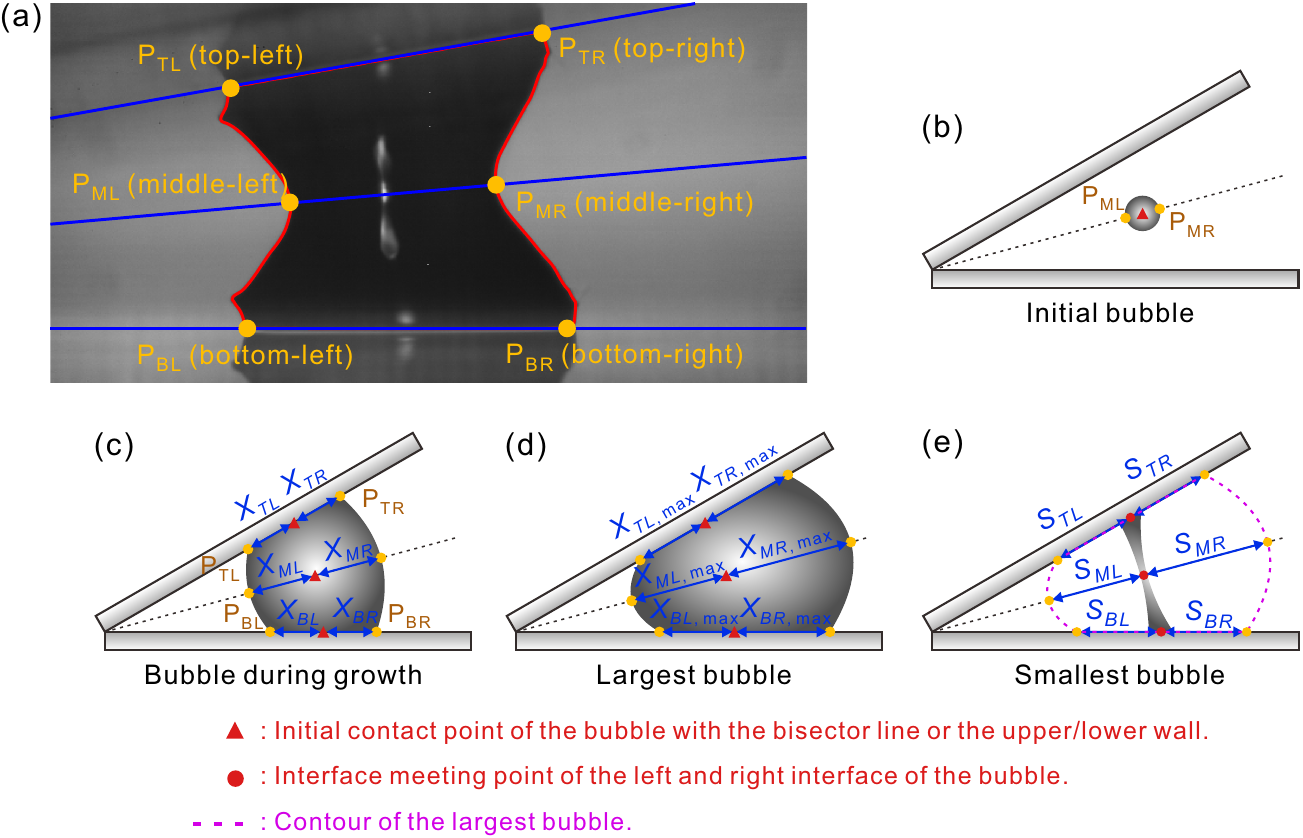}}
  \caption{Characterization of the bubble shape. (a) Definition of interaction points of the bubble contour with the upper wall, lower wall, and bisector line. (b-e) Definition of points, displacements, and shrinkage amounts at different instants of bubble evolution: (b) initial bubble, (c) bubble during growth, (d) largest bubble, and (d) smallest bubble. $X_{TL}$, $X_{TR}$, $X_{ML}$, $X_{MR}$, $X_{BL}$, and $X_BR$ are the displacements at the top/middle/bottom left/right side of the bubble. $S_{TL}$, $S_{TR}$, $S_{ML}$, $S_{MR}$, $S_{BL}$, and $S_{BR}$ are the shrinkage amounts at the top/middle/bottom left/right side of the bubble.
  $R_T = S_{TR}/S_{TL}$, $R_M = S_{MR}/S_{ML}$, and $R_B = S_{BR}/S_{BL}$ are the ratios of shrinkage between the left and right at the top/middle/bottom of the bubble.
  The first subscripts $T$, $M$ and $B$ indicate the top, middle, and bottom of the bubble.
  The second subscripts $L$ and $R$ indicate the Left and right interface of the bubble.
  }
\label{fig:10}
\end{figure}

To characterize the bubble shape and the interaction of the bubble with the wall, we performed digital image processing on the high-speed images and obtained the bubble contours from the high-speed images for every frame. Then we calculated the intersections of the contour curve with the upper wall, the lower wall, and the bisector line, respectively, as indicated by the six yellow dots in Figure \ref{fig:10}a. These six points are used to characterize the shape changes at the top, bottom, and middle of the bubble, respectively. Figure \ref{fig:10}(b-e) illustrates the shape changes of a bubble during the first cycle of bubble oscillation. The red triangles indicate the positions where the bubble first contacts with the upper wall, lower wall, and the bisector line. The initial contact point of the bubble with the bisector line is the midpoint of the line segment formed by points $\mathbf{P}_\text{ML}$ and $\mathbf{P}_\text{MR}$, when the bubble's contour first reaches the bisector. Similarly, the initial contact point of the bubble with the upper or lower wall is the midpoint of the line segment formed by points $\mathbf{P}_\text{TL}$ and $\mathbf{P}_\text{TR}$, or by points $\mathbf{P}_\text{LB}$ and $\mathbf{P}_\text{BR}$, when the bubble's contour first touches the upper or lower wall. Then, we calculated the displacement of the bubble at the top, middle, and bottom on both the left and right sides (see the length $X$ in Figure \ref{fig:10}). In addition, the interface meeting points (i.e., the point where the left and right sides of the bubble meet in the shrinkage process) on the top wall, on the bottom wall, or on the bisector line of the corner can be used to quantify the process of bubble evolution. Because the evolution of the bubble in the late stage of the contraction is extremely quick and challenging to capture accurately, we obtained the minimum shrinkage points by polynomial fitting. We fitted the points after the moments of maximum displacement on both the left and right sides of the bubble on the top wall, on the bottom wall, or on the bisector line of the corner, respectively, and the intersection points of the fitted curves for the left and right sides indicate the interface meeting points (i.e., the red dots in Figure \ref{fig:10}). Then, we calculated the shrinkage amount between the left and right sides (i.e., the length $S$ in Figure \ref{fig:10}). Here, the shrinkage amount means the distance between the maximum displacement point and the minimum shrinkage point. The ratio of shrinkage (i.e., $R$ in Figure \ref{fig:10}) between the left and right sides was also calculated, which enables a clear comparison of the difference between the shrinkage amounts on the left and right sides of the bubble.

\subsubsection{Effect of dimensionless lower-wall distance $d_{\mathrm{lower-wall}}^{ {*}}$}\label{sec:3.4.1}
\begin{figure}
  \centerline{\includegraphics[scale=0.75]{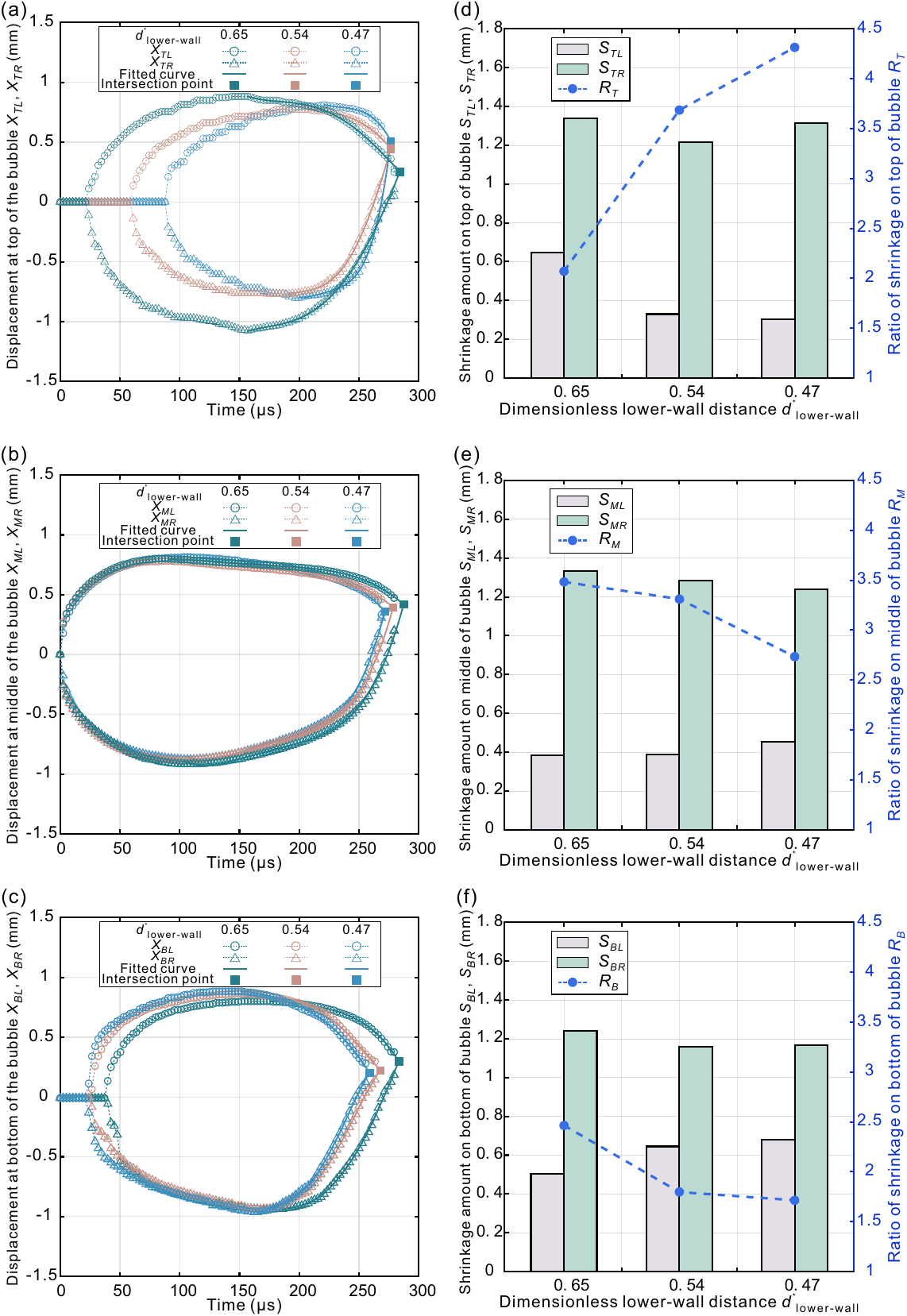}}
  \caption{(a-c) Temporal variation of the displacement of the top (a), middle (b), and bottom (c) of the bubble at a relatively small dimensionless vertex distance $d_{\text{vertex}}^{*} \approx 7.0$. The displacement is considered positive in the direction to the left. The circles represent the displacement of the bubble on the left side, whereas the triangles denote the displacement on the right side. Using polynomial fitting, we fitted all points after the moments of maximum displacement on both the left and right sides of the top, middle, and bottom of the bubble. The fitted curves are shown as solid lines. The intersection points of the fitted curves for the left and right sides are represented by solid squares. (d-f) Shrinkage amounts on both the left and right sides of the top (d), middle (e), and bottom (f) of the bubble at $d_{\text{vertex}}^{*} \approx 7.0$. Blue dots indicate the ratio of the right-side shrinkage to the left-side shrinkage.}
\label{fig:11}
\end{figure}

To quantify the bubble's dynamics, we monitored the evolution of the bubble shape during the bubble collapse. Figures \ref{fig:11}a, \ref{fig:11}b, and \ref{fig:11}c show the temporal variation of the displacement of the top, middle, and bottom of the bubble, respectively, at a relatively small dimensionless vertex distance $d_{\text{vertex}}^{*} \approx 7.0$, considering both the left and right sides. Since the top and bottom of the bubble do not contact the upper and lower walls initially during the early stage, the displacement values for both sides of the bubble are zero in the early stage in figures \ref{fig:11}a and \ref{fig:11}c.

Because of the asymmetric effect of the corner on the bubble dynamics, the top, middle, and bottom of the bubble exhibit asymmetry during collapse, as shown in Figure \ref{fig:11}. Regarding the top of the bubble (Figure \ref{fig:11}a), as the initial bubble position approaches the lower wall, there is a discernible difference in the slope of the displacement curve for the right side at different $d_{\text{lower-wall}}^{ {*}}$. At a relatively large dimensionless lower-wall distance (i.e., $d_{\text{lower-wall}}^{ {*}} = 0.65$), the displacement of the top right side of the bubble (i.e., the green solid line of $X_{TR}$ in Figure \ref{fig:11}a) exhibits relatively uniform changes throughout the shrinkage process. In contrast, as the dimensionless lower-wall distance decreases (i.e., $d_{\text{lower-wall}}^{ {*}} = 0.54$ and $d_{\text{lower-wall}}^{ {*}} = 0.47$), the displacement of the top right side of the bubble (i.e., the red and blue solid lines of $X_{TR}$ in Figure \ref{fig:11}a) shows minimal change in the early stages of shrinkage, resulting in a flatter displacement curve. However, the displacement curve suddenly becomes steep thereafter, indicating a sudden contraction of the top right side of the bubble. This is because, as the initial bubble position moves closer to the lower wall, the intensity of the annular shrinkage at the top of the bubble gradually diminishes, making the influence of the asymmetric structure of the corner on the top right side of the bubble more pronounced. This leads to a rapid contraction of the top right side, causing the displacement changes to become non-uniform. For the middle of the bubble (Figure \ref{fig:11}b), there remains a distinct asymmetry. The displacement variation of the right side of the middle of the bubble (i.e., $X_{MR}$ in Figure \ref{fig:11}b) is notably greater than that of the left side (i.e., $X_{ML}$ in Figure \ref{fig:11}b). Concerning the bottom of the bubble (Figure \ref{fig:11}c), when the dimensionless lower-wall distance is relatively large (i.e., $d_{\text{lower-wall}}^{ {*}} = 0.65$), there is already a strong annular shrinkage. As the bubble approaches the lower wall, this annular shrinkage continues to dominate, causing pronounced shrinkage on both the left and right sides of the bubble bottom. Thus, the asymmetry at the bubble bottom is less pronounced compared to that at the top and middle sections.

The shrinkage amounts on the left and right sides of the top of the bubble are compared in Figure \ref{fig:11}d. As $d_{\text{lower-wall}}^{ {*}}$ decreases, the initial position of the bubble progressively moves closer to the lower wall surface. Correspondingly, the difference in the shrinkage amounts between the left and right sides of the top of the bubble becomes increasingly pronounced, indicating that as the bubble moves downward, the asymmetry becomes more evident during the shrinking process. This is due to the increased distance between the top of the bubble and the upper wall, which weakens the annular shrinkage at the top of the bubble, consequently intensifying the asymmetry. The shrinkage amount on the right side of the top of the bubble is more than twice that of the left side. For the shrinkage amounts in the middle of the bubble (see Figure \ref{fig:11}e), under the influence of the corner structure, there is a considerable difference in the shrinkage amounts on the left and right sides as well. However, for the bottom of the bubble (see Figure \ref{fig:11}f), the difference in shrinkage amounts between its left and right sides is significantly smaller than that of the top and middle sections. Moreover, as the initial position of the bubble moves closer to the lower wall, the difference in shrinkage amounts between the left and right sides of the bubble bottom gradually diminishes. This is because, with the decrease in $d_{\text{lower-wall}}^{ {*}}$, the annular shrinkage at the bottom of the bubble intensifies, causing substantial shrinking on both the left and right sides, thereby reducing the asymmetry.

\begin{figure}
  \centerline{\includegraphics[scale=0.8]{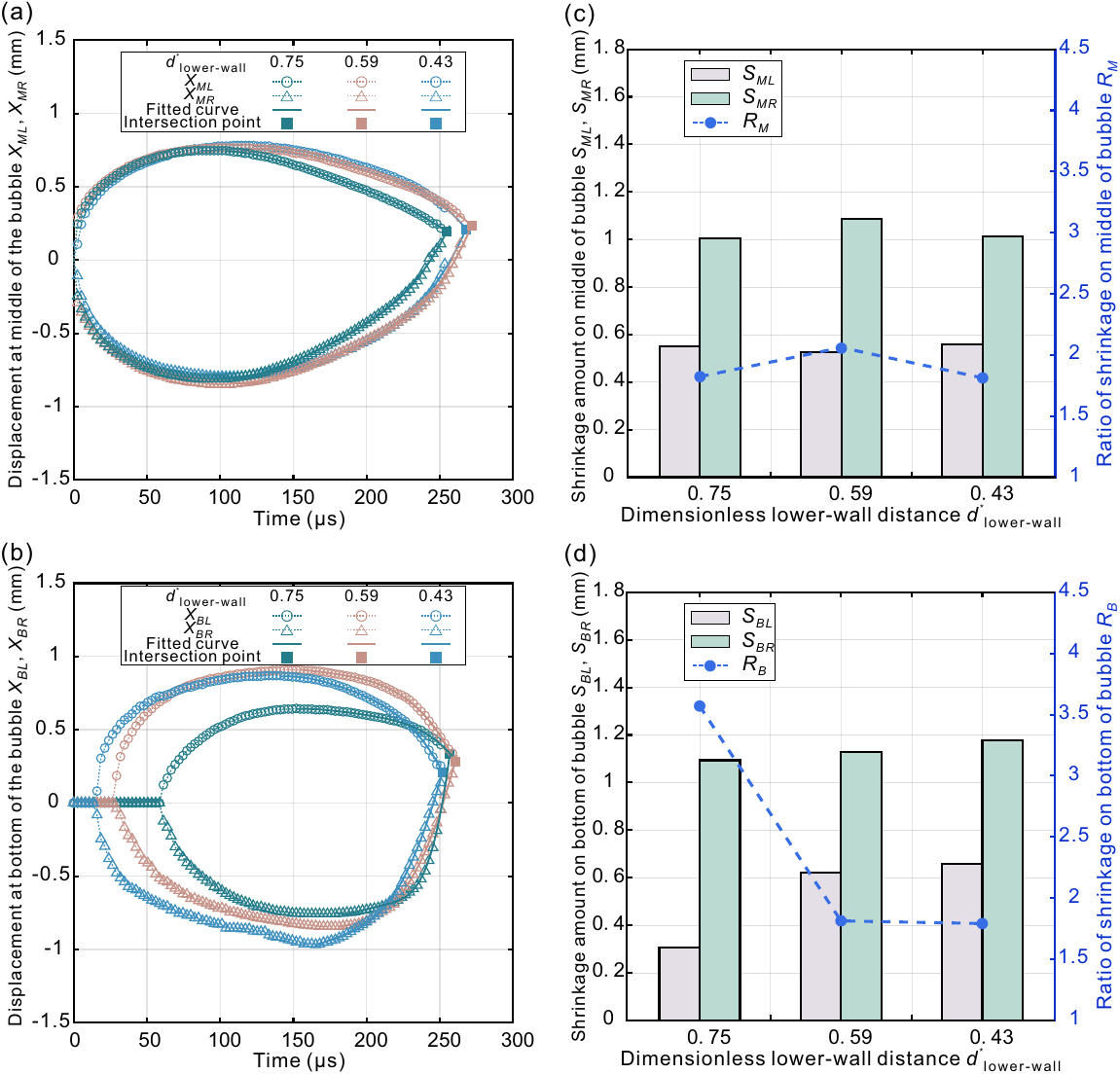}}
  \caption{(a-b) Time variation of displacement for the left and right sides of the bubble's middle (a) and bottom (b) at a moderate dimensionless vertex distance ($d_{\text{vertex}}^{*} \approx 8.3$). (c-d) Comparison of the shrinkage amounts for the bubble's middle (c) and bottom (d) at a moderate dimensionless vertex distance ($d_{\text{vertex}}^{*} \approx 8.3$).}
\label{fig:12}
\end{figure}

The variation of the displacement over time for the left and right sides of the bubble's middle and bottom, as well as a comparison of the shrinkage amounts for a moderate dimensionless vertex distance (i.e., $d_{\text{vertex}}^{*}\approx 8.3$) is shown in Figure \ref{fig:12}. When the dimensionless lower-wall distance is relatively small (i.e., $d_{\text{lower-wall}}^{ {*}} = 0.43$), the bubble is positioned far away from the upper wall, preventing its top from making contact with the upper wall (i.e., zero displacement for the left/right side of the top of the bubble). Therefore, we present the displacement for the middle and the bottom of the bubble only.

For the middle of the bubble, the asymmetry in the displacement curves for the left and right sides is less pronounced compared to that at a relatively small dimensionless vertex distance (i.e., $d_{\text{vertex}}^{*} \approx 7.0$), as shown in Figure \ref{fig:12}a. Furthermore, the shrinkage amount on the right side of the middle of the bubble is approximately twice that of the left side (Figure \ref{fig:12}c). Even as the initial position of the bubble gradually moves closer to the lower wall, the ratio of shrinkage amounts between the left and right sides exhibits no significant variation with $d_{\text{lower-wall}}^{ {*}}$. On the one hand, since the bubble at a moderate dimensionless vertex distance (i.e., $d_{\text{vertex}}^{*} \approx 8.3$) is situated farther from the vertex of the corner, it is subjected to less influence from the asymmetrical structure of the corner. On the other hand, the annular shrinkage in the middle of the bubble is more pronounced than that at a small dimensionless vertex distance (i.e., $d_{\text{vertex}}^{*} \approx 7.0$). As a result, both the left and right sides of the bubble experience a certain degree of shrinkage, which results in a less pronounced asymmetry.

For the bubble's bottom (Figure \ref{fig:12}b), when the dimensionless lower-wall distance is relatively large (i.e., $d_{\text{lower-wall}}^{ {*}} = 0.75$), there is virtually no annular shrinkage. Under the influence of the corner structure, unequal intensity of shrinkage occurs on the left and right sides of the bubble. The right side of the bottom of the bubble shrinks rapidly, leading to a very steep change in its displacement curve (see the green solid line of $X_{BR}$ in Figure \ref{fig:12}b). In contrast, the left side experiences minimal shrinks, resulting in a gradual change in its displacement curve (see the green solid line of $X_{BL}$ in Figure \ref{fig:12}b). This also gives rise to a considerable difference in the contraction amounts (as shown in Figure \ref{fig:12}d, $d_{\text{lower-wall}}^{ {*}} = 0.75$). As the dimensionless lower-wall distance decreases (i.e., $d_{\text{lower-wall}}^{ {*}} = 0.59$ and $d_{\text{lower-wall}}^{ {*}} = 0.43$), annular shrinkage occurs at the bottom of the bubble. Both the left and right sides of the bottom of the bubble undergo a certain degree of shrinkage, thus reducing the asymmetry. Ultimately, the shrinkage amount on the right side of the bubble bottom is approximately 1.5 times that of the left side (as shown in Figure \ref{fig:12}d, $d_{\text{lower-wall}}^{ {*}} = 0.59$ and $d_{\text{lower-wall}}^{ {*}} = 0.43$).

\subsubsection{Effect of dimensionless vertex distance $d_{\mathrm{vertex}}^{*}$}\label{sec:3.4.2}

\begin{figure}
  \centerline{\includegraphics[scale=0.8]{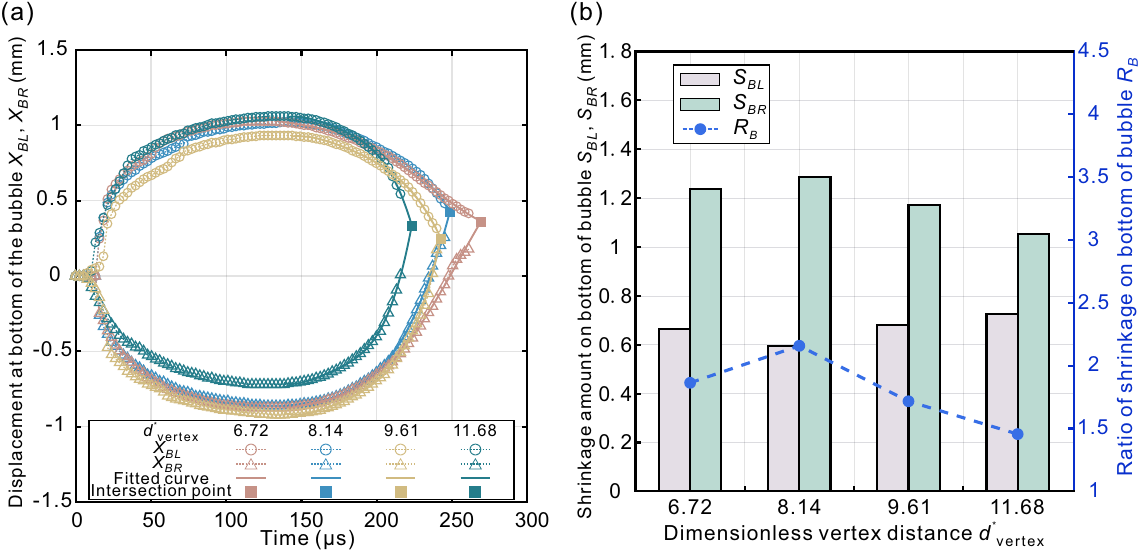}}
  \caption{(a) Displacement and (b) shrinkage amount comparisons for the left and right sides of the bottom of the bubble when the distance to the lower wall is fixed ($d^*_\text{lower-wall}=0.44$) and the bubble progressively moves away from the vertex of the corner.}
\label{fig:13}
\end{figure}
The displacement variations and shrinkage amount comparisons for the left and right sides of the bottom of the bubble when the bubble progressively moves away from the vertex of the corner are shown in Figure \ref{fig:13}. As the initial position of the bubble moves away from the vertex, the ratio of shrinkage amounts for the left and right sides of the bubble's bottom initially increases and then decreases. This is because, when the dimensionless vertex distance is relatively small (i.e., $d_{\text{vertex}}^{*} = 6.72$), the annular shrinkage at the bottom of the bubble is relatively intense. Both the left and right sides undergo a substantial degree of contraction, which reduces the asymmetry. As $d_{\text{vertex}}^{*}$ increases to 8.14, the annular shrinkage at the bottom of the bubble weakens. The asymmetry is more pronounced, leading to an increase in the ratio of shrinkage amounts. As $d_{\text{vertex}}^{*}$ continues to increase, the annular shrinkage at the bottom of the bubble further diminishes. However, as the initial position of the bubble moves farther away from the vertex of the corner, the influence of the corner on the bubble decreases. The bubble gradually transitions into symmetric contraction, causing the ratio of shrinkage amounts to decline, as shown in Figure \ref{fig:13}b.

\subsection{Bubble migration}\label{sec:3.5}
\begin{figure}
  \centerline{\includegraphics[scale=0.8]{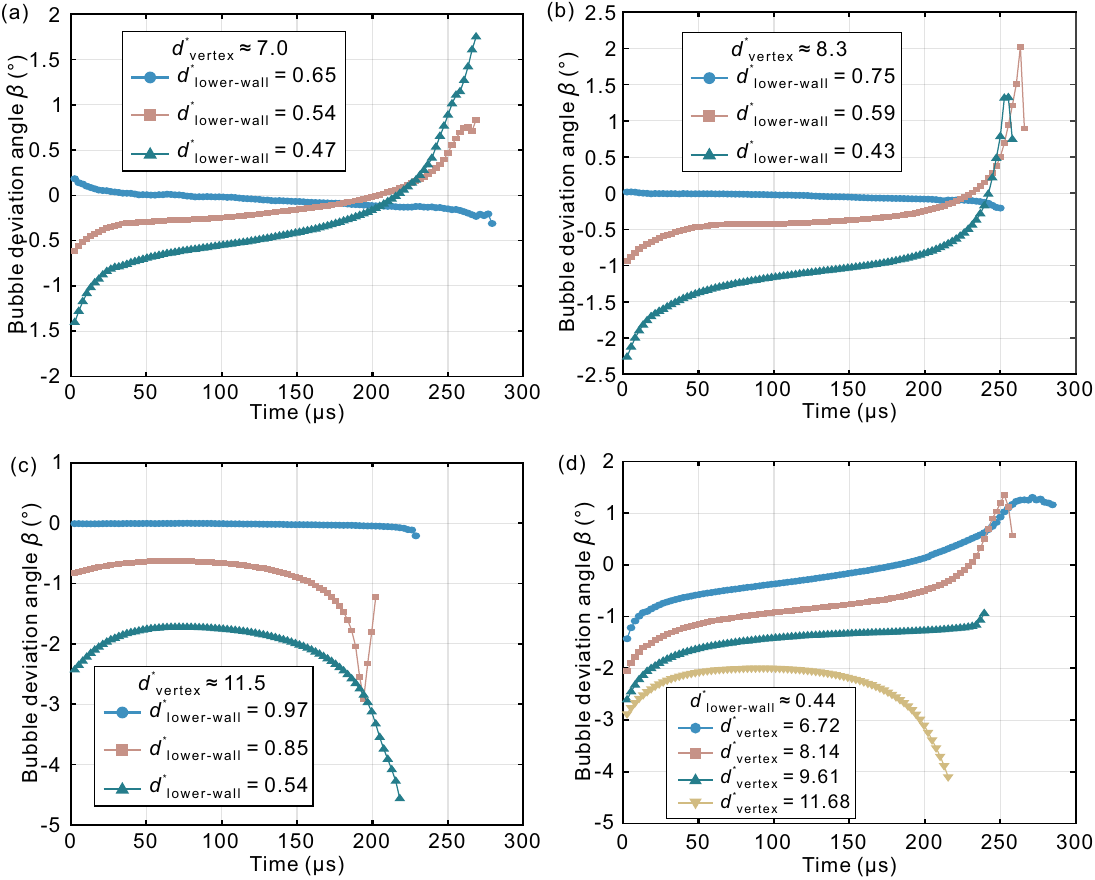}}
  \caption{Migration of the bubble at different initial bubble positions: (a) Bubbles generated very close to the vertex of the corner, with $d_{\text{vertex}}^{*} \approx 7.0$; (b) Bubbles generated relatively close to the vertex of the corner, with $d_{\text{vertex}}^{*} \approx 8.3$; (c) Bubbles generated relatively close to the vertex of the corner, with $d_{\text{vertex}}^{*} \approx 11.5$; (d) Bubbles generated with different dimensionless vertex distances with a fixed dimensionless lower-wall distance $d_{\text{lower-wall}}^{ {*}} \approx 0.44$. Here, the deviation angle of the bubble $\beta $ represents the angle formed by the line connecting the center of the bubble to the vertex of the corner and the bisector.}
\label{fig:14}
\end{figure}

To further quantify the overall position of the bubble approaching or departing from the wall over time, we consider the migration of the bubble, as shown in Figure \ref{fig:14}. Here, we use the deviation angle of the bubble $\beta $, which is the angle formed by the line connecting the center of the bubble to the vertex of the corner and the bisector. A value of $\beta = 0^\circ$ indicates that the bubble center is located on the bisector. As $\beta $ increases, it indicates that the bubble center moves closer to the upper wall, and vice versa.

For bubbles close to the vertex of the corner (see figures \ref{fig:14}a and \ref{fig:14}b), as the initial position of the bubble approaches the lower wall, the upward migration of the bubble center becomes increasingly pronounced. This is due to an increased intensity of annular shrinkage at the bottom of the bubble, causing the bubble to easily detach from the lower wall and migrate upward. However, it should be noted that at dimensionless lower-wall distances of $d_{\text{lower-wall}}^{ {*}} = 0.59$ and $d_{\text{lower-wall}}^{ {*}} = 0.43$, the bubble center undergoes downward migration during the final moments of collapse, as shown in Figure \ref{fig:14}b. This phenomenon is due to the progressive fragmentation of the bubble from bottom to top. When the lower part of the bubble breaks up, a large number of fine fragmented bubbles form, creating a large cluster of bubbles in the lower part of the image. Concurrently, the top part of the bubble is still undergoing shrinkage, resulting in a smaller bubble area at the upper half than at the lower half in the image. This leads to the apparent shift of the bubble center closer to the lower wall. When the bubble is far from the vertex (see $d_{\text{vertex}}^{*} \approx 11.5$ in Figure \ref{fig:14}c), as the position of the bubble approaches the lower wall, it undergoes downward migration rather than upward. However, at the dimensionless lower-wall distance $d_{\text{lower-wall}}^{ {*}} = 0.85$, during the final moments of collapse, the bubble center moves upward. This is, similarly, due to the fragmentation of the bubble from top to bottom. When a bubble is generated near the bisector (see $d_{\text{lower-wall}}^{ {*}} = 0.65$ in Figure \ref{fig:14}a, $d_{\text{lower-wall}}^{ {*}}= 0.75$ in Figure \ref{fig:14}b, and $d_{\text{lower-wall}}^{ {*}} = 0.97$ in Figure \ref{fig:14}c), the bubble center exhibits minimal displacement in the majority of the process. The bubble center experiences only a slight downward shift at the end of the collapse phase. This is because the initial bubble position may be slightly close to the upper wall. The shrinkage at the top of the bubble is slightly more pronounced, which can cause a slight downward displacement of the bubble during the collapse phase. When the bubble is moved further away from the vertex of the corner (i.e., increasing the dimensionless vertex distance $d_{\text{vertex}}^{*}$) while fixing the dimensionless lower-wall distance $d_{\text{lower-wall}}^{ {*}}$, the bubble center migration is shown in Figure \ref{fig:14}d. As the bubble moves progressively away from the corner vertex, the annular shrinkage at its bottom weakens. This reduces the upward migration of the bubble, leading to a transition of the bubble's motion from upward to downward.

\section{Conclusions}\label{sec:4}
In this study, the dynamics of cavitation bubbles in small corners are investigated experimentally, and the deformation and migration of the bubble are analyzed. The results show that the asymmetry of the corner leads to distinct scenarios of bubble shrinkage. The modes of bubble collapse can be divided mainly into three categories: collapse dominated by annular shrinkage, collapse dominated by wall attraction, and collapse governed by a combination of both annular shrinkage and wall attraction.

When the bubble is initiated very close to the vertex of the corner, the annular shrinkage dominates during the collapse process. As the initial position of the bubble moves to the lower wall, the annular shrinkage at the top of the bubble becomes weaker, while the annular shrinkage at the bottom of the bubble becomes stronger. This leads to a distinct asymmetry during the shrinkage of the left and right sides of the bubble top, whereas it is less pronounced at the bottom. Meanwhile, due to the strengthening of the annular shrinkage at the bottom, the upward migration of the bubble is more pronounced.

When the bubble is initiated far from the vertex of the corner, at the bisector of the angle, the bubble collapse is primarily dominated by a combination of annular shrinkage and wall attraction. However, in the vicinity of the wall, the bubble collapse is mainly governed by wall attraction. As the initial position of the bubble moves closer to the lower wall, there is no longer noticeable annular shrinkage at the bottom of the bubble. At this point, the attraction from the wall causes the bubble to either split or collapse downwards onto the lower wall. The center of the bubble gradually migrates downward.

When the initial position of the bubble progressively moves away from the vertex of the corner while the lower-wall distance is fixed, the asymmetry in the shrinkage at the bottom of the bubble initially increases and then decreases. This is due to the combined effect of the weakening annular shrinkage at the bottom of the bubble and the reduced asymmetric constraint imposed by the corner. Meanwhile, the bubble center shifts from migrating upwards to migrating downwards.

In summary, the asymmetric structure of the corner leads to asymmetric bubble dynamics, including asymmetric bubble expansion, spreading, contraction, and migration. The present results could be useful for applications of bubble cavitation, such as ultrasonic cleaning, micro-pumping, and underwater micro-propulsion. The damage to material surfaces induced by cavitation is a widespread phenomenon in fluid machinery and many other fields. Therefore, in the future, it will be crucial to study the mechanisms of surface damage caused by cavitation shock waves and jet impacts. Additionally, cavitation in other complex confined spaces is also important for the relevant applications and deserves more studies.

\section*{Acknowledgements}
This work is supported by the National Natural Science Foundation of China (Grant nos.\ 51920105010 and 51921004).

\section*{Data Availability Statement}
The data that support the findings of this study are available from the corresponding author upon reasonable request.

\section*{References}
\bibliography{BubbleCorner}

\begin{thebibliography}{46}%
\makeatletter
\providecommand \@ifxundefined [1]{%
 \@ifx{#1\undefined}
}%
\providecommand \@ifnum [1]{%
 \ifnum #1\expandafter \@firstoftwo
 \else \expandafter \@secondoftwo
 \fi
}%
\providecommand \@ifx [1]{%
 \ifx #1\expandafter \@firstoftwo
 \else \expandafter \@secondoftwo
 \fi
}%
\providecommand \natexlab [1]{#1}%
\providecommand \enquote  [1]{``#1''}%
\providecommand \bibnamefont  [1]{#1}%
\providecommand \bibfnamefont [1]{#1}%
\providecommand \citenamefont [1]{#1}%
\providecommand \href@noop [0]{\@secondoftwo}%
\providecommand \href [0]{\begingroup \@sanitize@url \@href}%
\providecommand \@href[1]{\@@startlink{#1}\@@href}%
\providecommand \@@href[1]{\endgroup#1\@@endlink}%
\providecommand \@sanitize@url [0]{\catcode `\\12\catcode `\$12\catcode
  `\&12\catcode `\#12\catcode `\^12\catcode `\_12\catcode `\%12\relax}%
\providecommand \@@startlink[1]{}%
\providecommand \@@endlink[0]{}%
\providecommand \url  [0]{\begingroup\@sanitize@url \@url }%
\providecommand \@url [1]{\endgroup\@href {#1}{\urlprefix }}%
\providecommand \urlprefix  [0]{URL }%
\providecommand \Eprint [0]{\href }%
\providecommand \doibase [0]{http://dx.doi.org/}%
\providecommand \selectlanguage [0]{\@gobble}%
\providecommand \bibinfo  [0]{\@secondoftwo}%
\providecommand \bibfield  [0]{\@secondoftwo}%
\providecommand \translation [1]{[#1]}%
\providecommand \BibitemOpen [0]{}%
\providecommand \bibitemStop [0]{}%
\providecommand \bibitemNoStop [0]{.\EOS\space}%
\providecommand \EOS [0]{\spacefactor3000\relax}%
\providecommand \BibitemShut  [1]{\csname bibitem#1\endcsname}%
\let\auto@bib@innerbib\@empty
\bibitem [{\citenamefont {Brennen}(2013)}]{brennen13}%
  \BibitemOpen
  \bibfield  {author} {\bibinfo {author} {\bibfnamefont {C.~E.}\ \bibnamefont
  {Brennen}},\ }\href {\doibase 10.1017/CBO9781107338760} {\emph {\bibinfo
  {title} {Cavitation and Bubble Dynamics}}}\ (\bibinfo  {publisher} {Cambridge
  University Press},\ \bibinfo {address} {Cambridge},\ \bibinfo {year}
  {2013})\BibitemShut {NoStop}%
\bibitem [{\citenamefont {Liao}\ \emph {et~al.}(2010)\citenamefont {Liao},
  \citenamefont {Ju}, \citenamefont {Zhang}, \citenamefont {He}, \citenamefont
  {Zhang}, \citenamefont {Shen}, \citenamefont {Chen}, \citenamefont {Cheng},
  \citenamefont {Xu}, \citenamefont {Sugioka},\ and\ \citenamefont
  {Midorikawa}}]{liao10}%
  \BibitemOpen
  \bibfield  {author} {\bibinfo {author} {\bibfnamefont {Y.}~\bibnamefont
  {Liao}}, \bibinfo {author} {\bibfnamefont {Y.}~\bibnamefont {Ju}}, \bibinfo
  {author} {\bibfnamefont {L.}~\bibnamefont {Zhang}}, \bibinfo {author}
  {\bibfnamefont {F.}~\bibnamefont {He}}, \bibinfo {author} {\bibfnamefont
  {Q.}~\bibnamefont {Zhang}}, \bibinfo {author} {\bibfnamefont
  {Y.}~\bibnamefont {Shen}}, \bibinfo {author} {\bibfnamefont {D.}~\bibnamefont
  {Chen}}, \bibinfo {author} {\bibfnamefont {Y.}~\bibnamefont {Cheng}},
  \bibinfo {author} {\bibfnamefont {Z.}~\bibnamefont {Xu}}, \bibinfo {author}
  {\bibfnamefont {K.}~\bibnamefont {Sugioka}}, \ and\ \bibinfo {author}
  {\bibfnamefont {K.}~\bibnamefont {Midorikawa}},\ }\bibfield  {title}
  {\enquote {\bibinfo {title} {Three-dimensional microfluidic channel with
  arbitrary length and configuration fabricated inside glass by femtosecond
  laser direct writing},}\ }\href {\doibase 10.1364/OL.35.003225} {\bibfield
  {journal} {\bibinfo  {journal} {Optics Letters}\ }\textbf {\bibinfo {volume}
  {35}},\ \bibinfo {pages} {3225--7} (\bibinfo {year} {2010})}\BibitemShut
  {NoStop}%
\bibitem [{\citenamefont {Shervani-Tabar}, \citenamefont {Abdullah},\ and\
  \citenamefont {Shabgard}(2006)}]{shervanitabar06}%
  \BibitemOpen
  \bibfield  {author} {\bibinfo {author} {\bibfnamefont {M.~T.}\ \bibnamefont
  {Shervani-Tabar}}, \bibinfo {author} {\bibfnamefont {A.}~\bibnamefont
  {Abdullah}}, \ and\ \bibinfo {author} {\bibfnamefont {M.~R.}\ \bibnamefont
  {Shabgard}},\ }\bibfield  {title} {\enquote {\bibinfo {title} {Numerical
  study on the dynamics of an electrical discharge generated bubble in
  {EDM}},}\ }\href {\doibase 10.1016/j.enganabound.2006.01.014} {\bibfield
  {journal} {\bibinfo  {journal} {Engineering Analysis with Boundary Elements}\
  }\textbf {\bibinfo {volume} {30}},\ \bibinfo {pages} {503--514} (\bibinfo
  {year} {2006})}\BibitemShut {NoStop}%
\bibitem [{\citenamefont {Hsiao}\ \emph {et~al.}(2013)\citenamefont {Hsiao},
  \citenamefont {Choi}, \citenamefont {Singh}, \citenamefont {Chahine},
  \citenamefont {Hay}, \citenamefont {Ilinskii}, \citenamefont {Zabolotskaya},
  \citenamefont {Hamilton}, \citenamefont {Sankin}, \citenamefont {Yuan},\ and\
  \citenamefont {Zhong}}]{hsiao13}%
  \BibitemOpen
  \bibfield  {author} {\bibinfo {author} {\bibfnamefont {C.~T.}\ \bibnamefont
  {Hsiao}}, \bibinfo {author} {\bibfnamefont {J.~K.}\ \bibnamefont {Choi}},
  \bibinfo {author} {\bibfnamefont {S.}~\bibnamefont {Singh}}, \bibinfo
  {author} {\bibfnamefont {G.~L.}\ \bibnamefont {Chahine}}, \bibinfo {author}
  {\bibfnamefont {T.~A.}\ \bibnamefont {Hay}}, \bibinfo {author} {\bibfnamefont
  {Y.~A.}\ \bibnamefont {Ilinskii}}, \bibinfo {author} {\bibfnamefont {E.~A.}\
  \bibnamefont {Zabolotskaya}}, \bibinfo {author} {\bibfnamefont {M.~F.}\
  \bibnamefont {Hamilton}}, \bibinfo {author} {\bibfnamefont {G.}~\bibnamefont
  {Sankin}}, \bibinfo {author} {\bibfnamefont {F.}~\bibnamefont {Yuan}}, \ and\
  \bibinfo {author} {\bibfnamefont {P.}~\bibnamefont {Zhong}},\ }\bibfield
  {title} {\enquote {\bibinfo {title} {Modelling single- and tandem-bubble
  dynamics between two parallel plates for biomedical applications},}\ }\href
  {\doibase 10.1017/jfm.2012.526} {\bibfield  {journal} {\bibinfo  {journal}
  {Journal of Fluid Mechanics}\ }\textbf {\bibinfo {volume} {716}},\ \bibinfo
  {pages} {137 -- 170} (\bibinfo {year} {2013})}\BibitemShut {NoStop}%
\bibitem [{\citenamefont {Mohammadzadeh}, \citenamefont {Li},\ and\
  \citenamefont {Ohl}(2017)}]{mohammadzadeh17}%
  \BibitemOpen
  \bibfield  {author} {\bibinfo {author} {\bibfnamefont {M.}~\bibnamefont
  {Mohammadzadeh}}, \bibinfo {author} {\bibfnamefont {F.}~\bibnamefont {Li}}, \
  and\ \bibinfo {author} {\bibfnamefont {C.-D.}\ \bibnamefont {Ohl}},\
  }\bibfield  {title} {\enquote {\bibinfo {title} {Shearing flow from transient
  bubble oscillations in narrow gaps},}\ }\href {\doibase
  10.1103/PhysRevFluids.2.014301} {\bibfield  {journal} {\bibinfo  {journal}
  {Physical Review Fluids}\ }\textbf {\bibinfo {volume} {2}},\ \bibinfo {pages}
  {014301} (\bibinfo {year} {2017})}\BibitemShut {NoStop}%
\bibitem [{\citenamefont {Yuan}, \citenamefont {Yang},\ and\ \citenamefont
  {Zhong}(2015)}]{yuan15}%
  \BibitemOpen
  \bibfield  {author} {\bibinfo {author} {\bibfnamefont {F.}~\bibnamefont
  {Yuan}}, \bibinfo {author} {\bibfnamefont {C.}~\bibnamefont {Yang}}, \ and\
  \bibinfo {author} {\bibfnamefont {P.}~\bibnamefont {Zhong}},\ }\bibfield
  {title} {\enquote {\bibinfo {title} {Cell membrane deformation and bioeffects
  produced by tandem bubble-induced jetting flow},}\ }\href {\doibase
  10.1073/pnas.1518679112} {\bibfield  {journal} {\bibinfo  {journal}
  {Proceedings of the National Academy of Sciences of the United States of
  America}\ }\textbf {\bibinfo {volume} {112}},\ \bibinfo {pages}
  {E7039--E7047} (\bibinfo {year} {2015})}\BibitemShut {NoStop}%
\bibitem [{\citenamefont {Blake}\ and\ \citenamefont {Cerone}(1982)}]{blake82}%
  \BibitemOpen
  \bibfield  {author} {\bibinfo {author} {\bibfnamefont {J.~R.}\ \bibnamefont
  {Blake}}\ and\ \bibinfo {author} {\bibfnamefont {P.}~\bibnamefont {Cerone}},\
  }\bibfield  {title} {\enquote {\bibinfo {title} {A note on the impulse due to
  a vapour bubble near a boundary},}\ }\href {\doibase
  10.1017/S0334270000000321} {\bibfield  {journal} {\bibinfo  {journal} {The
  Journal of the Australian Mathematical Society. Series B. Applied
  Mathematics}\ }\textbf {\bibinfo {volume} {23}},\ \bibinfo {pages} {383--393}
  (\bibinfo {year} {1982})}\BibitemShut {NoStop}%
\bibitem [{\citenamefont {Blake}\ and\ \citenamefont {Gibson}(1987)}]{blake87}%
  \BibitemOpen
  \bibfield  {author} {\bibinfo {author} {\bibfnamefont {J.~R.}\ \bibnamefont
  {Blake}}\ and\ \bibinfo {author} {\bibfnamefont {D.~C.}\ \bibnamefont
  {Gibson}},\ }\bibfield  {title} {\enquote {\bibinfo {title} {Cavitation
  bubbles near boundaries},}\ }\href {\doibase
  10.1146/annurev.fl.19.010187.000531} {\bibfield  {journal} {\bibinfo
  {journal} {Annual Review of Fluid Mechanics}\ }\textbf {\bibinfo {volume}
  {19}},\ \bibinfo {pages} {99--123} (\bibinfo {year} {1987})}\BibitemShut
  {NoStop}%
\bibitem [{\citenamefont {Chapman}\ and\ \citenamefont
  {Plesset}(1971)}]{chapman71}%
  \BibitemOpen
  \bibfield  {author} {\bibinfo {author} {\bibfnamefont {R.~B.}\ \bibnamefont
  {Chapman}}\ and\ \bibinfo {author} {\bibfnamefont {M.~S.}\ \bibnamefont
  {Plesset}},\ }\bibfield  {title} {\enquote {\bibinfo {title} {Collapse of an
  initially spherical vapour cavity in the neighbourhood of a solid
  boundary},}\ }\href {\doibase 10.1017/S0022112071001058} {\bibfield
  {journal} {\bibinfo  {journal} {Journal of Fluid Mechanics}\ }\textbf
  {\bibinfo {volume} {47}},\ \bibinfo {pages} {283--290} (\bibinfo {year}
  {1971})}\BibitemShut {NoStop}%
\bibitem [{\citenamefont {Dorsaz}\ \emph {et~al.}(2016)\citenamefont {Dorsaz},
  \citenamefont {Farhat}, \citenamefont {Kobel}, \citenamefont {Obreschkow},
  \citenamefont {Supponen},\ and\ \citenamefont {Tinguely}}]{dorsaz16}%
  \BibitemOpen
  \bibfield  {author} {\bibinfo {author} {\bibfnamefont {N.}~\bibnamefont
  {Dorsaz}}, \bibinfo {author} {\bibfnamefont {M.}~\bibnamefont {Farhat}},
  \bibinfo {author} {\bibfnamefont {P.}~\bibnamefont {Kobel}}, \bibinfo
  {author} {\bibfnamefont {D.}~\bibnamefont {Obreschkow}}, \bibinfo {author}
  {\bibfnamefont {O.}~\bibnamefont {Supponen}}, \ and\ \bibinfo {author}
  {\bibfnamefont {M.}~\bibnamefont {Tinguely}},\ }\bibfield  {title} {\enquote
  {\bibinfo {title} {Scaling laws for jets of single cavitation bubbles},}\
  }\href {\doibase 10.1017/jfm.2016.463} {\bibfield  {journal} {\bibinfo
  {journal} {Journal of Fluid Mechanics}\ }\textbf {\bibinfo {volume} {802}},\
  \bibinfo {pages} {263--293} (\bibinfo {year} {2016})}\BibitemShut {NoStop}%
\bibitem [{\citenamefont {Wang}\ and\ \citenamefont
  {Brennen}(1994)}]{wang1994shock}%
  \BibitemOpen
  \bibfield  {author} {\bibinfo {author} {\bibfnamefont {Y.-C.}\ \bibnamefont
  {Wang}}\ and\ \bibinfo {author} {\bibfnamefont {C.~E.}\ \bibnamefont
  {Brennen}},\ }\href@noop {} {\enquote {\bibinfo {title} {Shock wave
  development in the collapse of a cloud of bubbles},}\ }\bibinfo {type} {Tech.
  Rep.}\ (\bibinfo  {institution} {American Society of Mechanical Engineers,
  New York, NY (United States)},\ \bibinfo {year} {1994})\BibitemShut {NoStop}%
\bibitem [{\citenamefont {Gong}\ \emph {et~al.}(2024)\citenamefont {Gong},
  \citenamefont {Zhu}, \citenamefont {Ye},\ and\ \citenamefont
  {Fu}}]{gong2024numerical}%
  \BibitemOpen
  \bibfield  {author} {\bibinfo {author} {\bibfnamefont {T.}~\bibnamefont
  {Gong}}, \bibinfo {author} {\bibfnamefont {X.}~\bibnamefont {Zhu}}, \bibinfo
  {author} {\bibfnamefont {L.}~\bibnamefont {Ye}}, \ and\ \bibinfo {author}
  {\bibfnamefont {Y.}~\bibnamefont {Fu}},\ }\bibfield  {title} {\enquote
  {\bibinfo {title} {Numerical study of cavitation shock wave emission in the
  thin liquid layer by power ultrasonic vibratory machining},}\ }\href@noop {}
  {\bibfield  {journal} {\bibinfo  {journal} {Scientific Reports}\ }\textbf
  {\bibinfo {volume} {14}},\ \bibinfo {pages} {16956} (\bibinfo {year}
  {2024})}\BibitemShut {NoStop}%
\bibitem [{\citenamefont {Philipp}\ and\ \citenamefont
  {Lauterborn}(1998)}]{philipp1998cavitation}%
  \BibitemOpen
  \bibfield  {author} {\bibinfo {author} {\bibfnamefont {A.}~\bibnamefont
  {Philipp}}\ and\ \bibinfo {author} {\bibfnamefont {W.}~\bibnamefont
  {Lauterborn}},\ }\bibfield  {title} {\enquote {\bibinfo {title} {Cavitation
  erosion by single laser-produced bubbles},}\ }\href@noop {} {\bibfield
  {journal} {\bibinfo  {journal} {Journal of fluid mechanics}\ }\textbf
  {\bibinfo {volume} {361}},\ \bibinfo {pages} {75--116} (\bibinfo {year}
  {1998})}\BibitemShut {NoStop}%
\bibitem [{\citenamefont {Lauterborn}\ and\ \citenamefont
  {Philipp}(1998)}]{lauterborn98}%
  \BibitemOpen
  \bibfield  {author} {\bibinfo {author} {\bibfnamefont {W.}~\bibnamefont
  {Lauterborn}}\ and\ \bibinfo {author} {\bibfnamefont {A.}~\bibnamefont
  {Philipp}},\ }\bibfield  {title} {\enquote {\bibinfo {title} {Cavitation
  erosion by single laser-produced bubbles},}\ }\href {\doibase
  10.1017/S0022112098008738} {\bibfield  {journal} {\bibinfo  {journal}
  {Journal of Fluid Mechanics}\ }\textbf {\bibinfo {volume} {361}},\ \bibinfo
  {pages} {75--116} (\bibinfo {year} {1998})}\BibitemShut {NoStop}%
\bibitem [{\citenamefont {Luo}\ \emph {et~al.}(2018)\citenamefont {Luo},
  \citenamefont {Xu}, \citenamefont {Deng}, \citenamefont {Zhai},\ and\
  \citenamefont {Zhang}}]{luo18}%
  \BibitemOpen
  \bibfield  {author} {\bibinfo {author} {\bibfnamefont {J.}~\bibnamefont
  {Luo}}, \bibinfo {author} {\bibfnamefont {W.}~\bibnamefont {Xu}}, \bibinfo
  {author} {\bibfnamefont {J.}~\bibnamefont {Deng}}, \bibinfo {author}
  {\bibfnamefont {Y.}~\bibnamefont {Zhai}}, \ and\ \bibinfo {author}
  {\bibfnamefont {Q.}~\bibnamefont {Zhang}},\ }\bibfield  {title} {\enquote
  {\bibinfo {title} {Experimental study on the impact characteristics of
  cavitation bubble collapse on a wall},}\ }\href {\doibase 10.3390/w10091262}
  {\bibfield  {journal} {\bibinfo  {journal} {Water}\ }\textbf {\bibinfo
  {volume} {10}},\ \bibinfo {pages} {104951} (\bibinfo {year}
  {2018})}\BibitemShut {NoStop}%
\bibitem [{\citenamefont {Lechner}\ \emph {et~al.}(2020)\citenamefont
  {Lechner}, \citenamefont {Lauterborn}, \citenamefont {Koch},\ and\
  \citenamefont {Mettin}}]{lechner20}%
  \BibitemOpen
  \bibfield  {author} {\bibinfo {author} {\bibfnamefont {C.}~\bibnamefont
  {Lechner}}, \bibinfo {author} {\bibfnamefont {W.}~\bibnamefont {Lauterborn}},
  \bibinfo {author} {\bibfnamefont {M.}~\bibnamefont {Koch}}, \ and\ \bibinfo
  {author} {\bibfnamefont {R.}~\bibnamefont {Mettin}},\ }\bibfield  {title}
  {\enquote {\bibinfo {title} {Jet formation from bubbles near a solid boundary
  in a compressible liquid: Numerical study of distance dependence},}\ }\href
  {\doibase 10.1103/PhysRevFluids.5.093604} {\bibfield  {journal} {\bibinfo
  {journal} {Physical Review Fluids}\ }\textbf {\bibinfo {volume} {5}},\
  \bibinfo {pages} {093604} (\bibinfo {year} {2020})}\BibitemShut {NoStop}%
\bibitem [{\citenamefont {Dijkink}\ and\ \citenamefont
  {Ohl}(2008)}]{dijkink08}%
  \BibitemOpen
  \bibfield  {author} {\bibinfo {author} {\bibfnamefont {R.}~\bibnamefont
  {Dijkink}}\ and\ \bibinfo {author} {\bibfnamefont {C.-D.}\ \bibnamefont
  {Ohl}},\ }\bibfield  {title} {\enquote {\bibinfo {title} {Measurement of
  cavitation induced wall shear stress},}\ }\href {\doibase 10.1063/1.3046735}
  {\bibfield  {journal} {\bibinfo  {journal} {Applied Physics Letters}\
  }\textbf {\bibinfo {volume} {93}},\ \bibinfo {pages} {254107} (\bibinfo
  {year} {2008})}\BibitemShut {NoStop}%
\bibitem [{\citenamefont {Occhicone}\ \emph {et~al.}(2019)\citenamefont
  {Occhicone}, \citenamefont {Sinibaldi}, \citenamefont {Danz}, \citenamefont
  {Casciola},\ and\ \citenamefont {Michelotti}}]{occhicone19}%
  \BibitemOpen
  \bibfield  {author} {\bibinfo {author} {\bibfnamefont {A.}~\bibnamefont
  {Occhicone}}, \bibinfo {author} {\bibfnamefont {G.}~\bibnamefont
  {Sinibaldi}}, \bibinfo {author} {\bibfnamefont {N.}~\bibnamefont {Danz}},
  \bibinfo {author} {\bibfnamefont {C.~M.}\ \bibnamefont {Casciola}}, \ and\
  \bibinfo {author} {\bibfnamefont {F.}~\bibnamefont {Michelotti}},\ }\bibfield
   {title} {\enquote {\bibinfo {title} {Cavitation bubble wall pressure
  measurement by an electromagnetic surface wave enhanced pump-probe
  configuration},}\ }\href {\doibase 10.1063/1.5089206} {\bibfield  {journal}
  {\bibinfo  {journal} {Applied Physics Letters}\ }\textbf {\bibinfo {volume}
  {114}},\ \bibinfo {pages} {134101} (\bibinfo {year} {2019})}\BibitemShut
  {NoStop}%
\bibitem [{\citenamefont {Dijkink}\ \emph {et~al.}(2018)\citenamefont
  {Dijkink}, \citenamefont {Gavaises}, \citenamefont {Gonzalez-Avila},
  \citenamefont {Koukouvinis}, \citenamefont {Ohl},\ and\ \citenamefont
  {Zeng}}]{dijkink18}%
  \BibitemOpen
  \bibfield  {author} {\bibinfo {author} {\bibfnamefont {R.}~\bibnamefont
  {Dijkink}}, \bibinfo {author} {\bibfnamefont {M.}~\bibnamefont {Gavaises}},
  \bibinfo {author} {\bibfnamefont {S.~R.}\ \bibnamefont {Gonzalez-Avila}},
  \bibinfo {author} {\bibfnamefont {P.}~\bibnamefont {Koukouvinis}}, \bibinfo
  {author} {\bibfnamefont {C.-D.}\ \bibnamefont {Ohl}}, \ and\ \bibinfo
  {author} {\bibfnamefont {Q.}~\bibnamefont {Zeng}},\ }\bibfield  {title}
  {\enquote {\bibinfo {title} {Wall shear stress from jetting cavitation
  bubbles},}\ }\href {\doibase 10.1017/jfm.2018.286} {\bibfield  {journal}
  {\bibinfo  {journal} {Journal of Fluid Mechanics}\ }\textbf {\bibinfo
  {volume} {846}},\ \bibinfo {pages} {341--355} (\bibinfo {year}
  {2018})}\BibitemShut {NoStop}%
\bibitem [{\citenamefont {Li}\ \emph {et~al.}(2016)\citenamefont {Li},
  \citenamefont {Han}, \citenamefont {Zhang},\ and\ \citenamefont
  {Wang}}]{li16}%
  \BibitemOpen
  \bibfield  {author} {\bibinfo {author} {\bibfnamefont {S.}~\bibnamefont
  {Li}}, \bibinfo {author} {\bibfnamefont {R.}~\bibnamefont {Han}}, \bibinfo
  {author} {\bibfnamefont {A.~M.}\ \bibnamefont {Zhang}}, \ and\ \bibinfo
  {author} {\bibfnamefont {Q.~X.}\ \bibnamefont {Wang}},\ }\bibfield  {title}
  {\enquote {\bibinfo {title} {Analysis of pressure field generated by a
  collapsing bubble},}\ }\href {\doibase 10.1016/j.oceaneng.2016.03.016}
  {\bibfield  {journal} {\bibinfo  {journal} {Ocean Engineering}\ }\textbf
  {\bibinfo {volume} {117}},\ \bibinfo {pages} {22--38} (\bibinfo {year}
  {2016})}\BibitemShut {NoStop}%
\bibitem [{\citenamefont {Gonzalez-Avila}, \citenamefont {Ohl},\ and\
  \citenamefont {Zeng}(2020)}]{gonzalezavila20}%
  \BibitemOpen
  \bibfield  {author} {\bibinfo {author} {\bibfnamefont {S.~R.}\ \bibnamefont
  {Gonzalez-Avila}}, \bibinfo {author} {\bibfnamefont {C.-D.}\ \bibnamefont
  {Ohl}}, \ and\ \bibinfo {author} {\bibfnamefont {Q.}~\bibnamefont {Zeng}},\
  }\bibfield  {title} {\enquote {\bibinfo {title} {Splitting and jetting of
  cavitation bubbles in thin gaps},}\ }\href {\doibase 10.1017/jfm.2020.356}
  {\bibfield  {journal} {\bibinfo  {journal} {Journal of Fluid Mechanics}\
  }\textbf {\bibinfo {volume} {896}},\ \bibinfo {pages} {A28} (\bibinfo {year}
  {2020})}\BibitemShut {NoStop}%
\bibitem [{\citenamefont {Quinto-Su}, \citenamefont {Lim},\ and\ \citenamefont
  {Ohl}(2009)}]{quinto-su09}%
  \BibitemOpen
  \bibfield  {author} {\bibinfo {author} {\bibfnamefont {P.~A.}\ \bibnamefont
  {Quinto-Su}}, \bibinfo {author} {\bibfnamefont {K.~Y.}\ \bibnamefont {Lim}},
  \ and\ \bibinfo {author} {\bibfnamefont {C.-D.}\ \bibnamefont {Ohl}},\
  }\bibfield  {title} {\enquote {\bibinfo {title} {Cavitation bubble dynamics
  in microfluidic gaps of variable height},}\ }\href {\doibase
  10.1103/PhysRevE.80.047301} {\bibfield  {journal} {\bibinfo  {journal}
  {Physical Review E}\ }\textbf {\bibinfo {volume} {80}},\ \bibinfo {pages}
  {047301} (\bibinfo {year} {2009})}\BibitemShut {NoStop}%
\bibitem [{\citenamefont {Gonzalez-Avila}\ \emph {et~al.}(2011)\citenamefont
  {Gonzalez-Avila}, \citenamefont {Khoo}, \citenamefont {Klaseboer},\ and\
  \citenamefont {Ohl}}]{gonzalezavila11}%
  \BibitemOpen
  \bibfield  {author} {\bibinfo {author} {\bibfnamefont {S.~R.}\ \bibnamefont
  {Gonzalez-Avila}}, \bibinfo {author} {\bibfnamefont {B.~C.}\ \bibnamefont
  {Khoo}}, \bibinfo {author} {\bibfnamefont {E.}~\bibnamefont {Klaseboer}}, \
  and\ \bibinfo {author} {\bibfnamefont {C.-D.}\ \bibnamefont {Ohl}},\
  }\bibfield  {title} {\enquote {\bibinfo {title} {Cavitation bubble dynamics
  in a liquid gap of variable height},}\ }\href {\doibase 10.1017/jfm.2011.212}
  {\bibfield  {journal} {\bibinfo  {journal} {Journal of Fluid Mechanics}\
  }\textbf {\bibinfo {volume} {682}},\ \bibinfo {pages} {241--260} (\bibinfo
  {year} {2011})}\BibitemShut {NoStop}%
\bibitem [{\citenamefont {Brujan}, \citenamefont {Takahira},\ and\
  \citenamefont {Ogasawara}(2019)}]{brujan19}%
  \BibitemOpen
  \bibfield  {author} {\bibinfo {author} {\bibfnamefont {E.-A.}\ \bibnamefont
  {Brujan}}, \bibinfo {author} {\bibfnamefont {H.}~\bibnamefont {Takahira}}, \
  and\ \bibinfo {author} {\bibfnamefont {T.}~\bibnamefont {Ogasawara}},\
  }\bibfield  {title} {\enquote {\bibinfo {title} {Planar jets in collapsing
  cavitation bubbles},}\ }\href {\doibase 10.1016/j.expthermflusci.2018.10.007}
  {\bibfield  {journal} {\bibinfo  {journal} {Experimental Thermal and Fluid
  Science}\ }\textbf {\bibinfo {volume} {101}},\ \bibinfo {pages} {48--61}
  (\bibinfo {year} {2019})}\BibitemShut {NoStop}%
\bibitem [{\citenamefont {Gonzalez-Avila}\ \emph {et~al.}(2020)\citenamefont
  {Gonzalez-Avila}, \citenamefont {van Blokland}, \citenamefont {Zeng},\ and\
  \citenamefont {Ohl}}]{gonzalezavila20Jetting}%
  \BibitemOpen
  \bibfield  {author} {\bibinfo {author} {\bibfnamefont {S.~R.}\ \bibnamefont
  {Gonzalez-Avila}}, \bibinfo {author} {\bibfnamefont {A.~C.}\ \bibnamefont
  {van Blokland}}, \bibinfo {author} {\bibfnamefont {Q.}~\bibnamefont {Zeng}},
  \ and\ \bibinfo {author} {\bibfnamefont {C.-D.}\ \bibnamefont {Ohl}},\
  }\bibfield  {title} {\enquote {\bibinfo {title} {Jetting and shear stress
  enhancement from cavitation bubbles collapsing in a narrow gap},}\ }\href
  {\doibase 10.1017/jfm.2019.938} {\bibfield  {journal} {\bibinfo  {journal}
  {Journal of Fluid Mechanics}\ }\textbf {\bibinfo {volume} {884}},\ \bibinfo
  {pages} {A23} (\bibinfo {year} {2020})}\BibitemShut {NoStop}%
\bibitem [{\citenamefont {Quah}\ \emph {et~al.}(2018)\citenamefont {Quah},
  \citenamefont {Karri}, \citenamefont {Ohl}, \citenamefont {Klaseboer},\ and\
  \citenamefont {Khoo}}]{quah18}%
  \BibitemOpen
  \bibfield  {author} {\bibinfo {author} {\bibfnamefont {E.~W.}\ \bibnamefont
  {Quah}}, \bibinfo {author} {\bibfnamefont {B.}~\bibnamefont {Karri}},
  \bibinfo {author} {\bibfnamefont {S.-W.}\ \bibnamefont {Ohl}}, \bibinfo
  {author} {\bibfnamefont {E.}~\bibnamefont {Klaseboer}}, \ and\ \bibinfo
  {author} {\bibfnamefont {B.~C.}\ \bibnamefont {Khoo}},\ }\bibfield  {title}
  {\enquote {\bibinfo {title} {Expansion and collapse of an initially
  off-centered bubble within a narrow gap and the effect of a free surface},}\
  }\href {\doibase 10.1016/j.ijmultiphaseflow.2017.09.013} {\bibfield
  {journal} {\bibinfo  {journal} {International Journal of Multiphase Flow}\
  }\textbf {\bibinfo {volume} {99}},\ \bibinfo {pages} {62--72} (\bibinfo
  {year} {2018})}\BibitemShut {NoStop}%
\bibitem [{\citenamefont {Cui}\ \emph {et~al.}(2015)\citenamefont {Cui},
  \citenamefont {Cui}, \citenamefont {Wang},\ and\ \citenamefont
  {Zhang}}]{cui15}%
  \BibitemOpen
  \bibfield  {author} {\bibinfo {author} {\bibfnamefont {P.}~\bibnamefont
  {Cui}}, \bibinfo {author} {\bibfnamefont {J.}~\bibnamefont {Cui}}, \bibinfo
  {author} {\bibfnamefont {Q.~X.}\ \bibnamefont {Wang}}, \ and\ \bibinfo
  {author} {\bibfnamefont {A.~M.}\ \bibnamefont {Zhang}},\ }\bibfield  {title}
  {\enquote {\bibinfo {title} {Experimental study on bubble dynamics subject to
  buoyancy},}\ }\href {\doibase 10.1017/jfm.2015.323} {\bibfield  {journal}
  {\bibinfo  {journal} {Journal of Fluid Mechanics}\ }\textbf {\bibinfo
  {volume} {776}},\ \bibinfo {pages} {137--160} (\bibinfo {year}
  {2015})}\BibitemShut {NoStop}%
\bibitem [{\citenamefont {Khoo}\ \emph {et~al.}(2009)\citenamefont {Khoo},
  \citenamefont {Adikhari}, \citenamefont {Fong},\ and\ \citenamefont
  {Klaseboer}}]{khoo09}%
  \BibitemOpen
  \bibfield  {author} {\bibinfo {author} {\bibfnamefont {B.~C.}\ \bibnamefont
  {Khoo}}, \bibinfo {author} {\bibfnamefont {D.}~\bibnamefont {Adikhari}},
  \bibinfo {author} {\bibfnamefont {S.~W.}\ \bibnamefont {Fong}}, \ and\
  \bibinfo {author} {\bibfnamefont {E.}~\bibnamefont {Klaseboer}},\ }\bibfield
  {title} {\enquote {\bibinfo {title} {Multiple spark-generated bubble
  interactions},}\ }\href {\doibase 10.1142/S0217984909018072} {\bibfield
  {journal} {\bibinfo  {journal} {Modern Physics Letters B}\ }\textbf {\bibinfo
  {volume} {23}},\ \bibinfo {pages} {229--232} (\bibinfo {year}
  {2009})}\BibitemShut {NoStop}%
\bibitem [{\citenamefont {Azam}\ \emph {et~al.}(2013)\citenamefont {Azam},
  \citenamefont {Karri}, \citenamefont {Ohl}, \citenamefont {Klaseboer},\ and\
  \citenamefont {Khoo}}]{azam13}%
  \BibitemOpen
  \bibfield  {author} {\bibinfo {author} {\bibfnamefont {F.~I.}\ \bibnamefont
  {Azam}}, \bibinfo {author} {\bibfnamefont {B.}~\bibnamefont {Karri}},
  \bibinfo {author} {\bibfnamefont {S.-W.}\ \bibnamefont {Ohl}}, \bibinfo
  {author} {\bibfnamefont {E.}~\bibnamefont {Klaseboer}}, \ and\ \bibinfo
  {author} {\bibfnamefont {B.~C.}\ \bibnamefont {Khoo}},\ }\bibfield  {title}
  {\enquote {\bibinfo {title} {Dynamics of an oscillating bubble in a narrow
  gap},}\ }\href {\doibase 10.1103/PhysRevE.88.043006} {\bibfield  {journal}
  {\bibinfo  {journal} {Physical Review E}\ }\textbf {\bibinfo {volume} {88}},\
  \bibinfo {pages} {043006} (\bibinfo {year} {2013})}\BibitemShut {NoStop}%
\bibitem [{\citenamefont {Sagar}\ and\ \citenamefont
  {El~Moctar}(2023)}]{sagar23}%
  \BibitemOpen
  \bibfield  {author} {\bibinfo {author} {\bibfnamefont {H.~J.}\ \bibnamefont
  {Sagar}}\ and\ \bibinfo {author} {\bibfnamefont {O.}~\bibnamefont
  {El~Moctar}},\ }\bibfield  {title} {\enquote {\bibinfo {title} {Dynamics of a
  cavitation bubble between oblique plates},}\ }\href {\doibase
  10.1063/5.0132098} {\bibfield  {journal} {\bibinfo  {journal} {Physics of
  Fluids}\ }\textbf {\bibinfo {volume} {35}},\ \bibinfo {pages} {013324}
  (\bibinfo {year} {2023})}\BibitemShut {NoStop}%
\bibitem [{\citenamefont {Brujan}\ \emph {et~al.}(2018)\citenamefont {Brujan},
  \citenamefont {Ishigami}, \citenamefont {Noda}, \citenamefont {Ogasawara},\
  and\ \citenamefont {Takahira}}]{brujan18}%
  \BibitemOpen
  \bibfield  {author} {\bibinfo {author} {\bibfnamefont {E.-A.}\ \bibnamefont
  {Brujan}}, \bibinfo {author} {\bibfnamefont {A.}~\bibnamefont {Ishigami}},
  \bibinfo {author} {\bibfnamefont {T.}~\bibnamefont {Noda}}, \bibinfo {author}
  {\bibfnamefont {T.}~\bibnamefont {Ogasawara}}, \ and\ \bibinfo {author}
  {\bibfnamefont {H.}~\bibnamefont {Takahira}},\ }\bibfield  {title} {\enquote
  {\bibinfo {title} {Dynamics of laser-induced cavitation bubbles near two
  perpendicular rigid walls},}\ }\href {\doibase 10.1017/jfm.2018.82}
  {\bibfield  {journal} {\bibinfo  {journal} {Journal of Fluid Mechanics}\
  }\textbf {\bibinfo {volume} {841}},\ \bibinfo {pages} {28--49} (\bibinfo
  {year} {2018})}\BibitemShut {NoStop}%
\bibitem [{\citenamefont {White}, \citenamefont {Beig},\ and\ \citenamefont
  {Johnsen}(2023)}]{white23}%
  \BibitemOpen
  \bibfield  {author} {\bibinfo {author} {\bibfnamefont {W.}~\bibnamefont
  {White}}, \bibinfo {author} {\bibfnamefont {S.~A.}\ \bibnamefont {Beig}}, \
  and\ \bibinfo {author} {\bibfnamefont {E.}~\bibnamefont {Johnsen}},\
  }\bibfield  {title} {\enquote {\bibinfo {title} {Pressure fields produced by
  single-bubble collapse near a corner},}\ }\href {\doibase
  10.1103/PhysRevFluids.8.023601} {\bibfield  {journal} {\bibinfo  {journal}
  {Physical Review Fluids}\ }\textbf {\bibinfo {volume} {8}},\ \bibinfo {pages}
  {023601} (\bibinfo {year} {2023})}\BibitemShut {NoStop}%
\bibitem [{\citenamefont {Wang}\ \emph {et~al.}(2020)\citenamefont {Wang},
  \citenamefont {Mahmud}, \citenamefont {Cui}, \citenamefont {Smith},\ and\
  \citenamefont {Walmsley}}]{wang20}%
  \BibitemOpen
  \bibfield  {author} {\bibinfo {author} {\bibfnamefont {Q.}~\bibnamefont
  {Wang}}, \bibinfo {author} {\bibfnamefont {M.}~\bibnamefont {Mahmud}},
  \bibinfo {author} {\bibfnamefont {J.}~\bibnamefont {Cui}}, \bibinfo {author}
  {\bibfnamefont {W.~R.}\ \bibnamefont {Smith}}, \ and\ \bibinfo {author}
  {\bibfnamefont {A.~D.}\ \bibnamefont {Walmsley}},\ }\bibfield  {title}
  {\enquote {\bibinfo {title} {Numerical investigation of bubble dynamics at a
  corner},}\ }\href {\doibase 10.1063/1.5140740} {\bibfield  {journal}
  {\bibinfo  {journal} {Physics of Fluids}\ }\textbf {\bibinfo {volume} {32}},\
  \bibinfo {pages} {053306} (\bibinfo {year} {2020})}\BibitemShut {NoStop}%
\bibitem [{\citenamefont {Cui}\ \emph {et~al.}(2020)\citenamefont {Cui},
  \citenamefont {Chen}, \citenamefont {Wang}, \citenamefont {Zhou},\ and\
  \citenamefont {Corbett}}]{cui20}%
  \BibitemOpen
  \bibfield  {author} {\bibinfo {author} {\bibfnamefont {J.}~\bibnamefont
  {Cui}}, \bibinfo {author} {\bibfnamefont {Z.-P.}\ \bibnamefont {Chen}},
  \bibinfo {author} {\bibfnamefont {Q.}~\bibnamefont {Wang}}, \bibinfo {author}
  {\bibfnamefont {T.-R.}\ \bibnamefont {Zhou}}, \ and\ \bibinfo {author}
  {\bibfnamefont {C.}~\bibnamefont {Corbett}},\ }\bibfield  {title} {\enquote
  {\bibinfo {title} {Experimental studies of bubble dynamics inside a
  corner},}\ }\href {\doibase 10.1016/j.ultsonch.2019.104951} {\bibfield
  {journal} {\bibinfo  {journal} {Ultrasonics Sonochemistry}\ }\textbf
  {\bibinfo {volume} {64}},\ \bibinfo {pages} {104951} (\bibinfo {year}
  {2020})}\BibitemShut {NoStop}%
\bibitem [{\citenamefont {Cui}\ \emph {et~al.}(2023)\citenamefont {Cui},
  \citenamefont {Li}, \citenamefont {Li}, \citenamefont {Liu},\ and\
  \citenamefont {Zhang}}]{cui23}%
  \BibitemOpen
  \bibfield  {author} {\bibinfo {author} {\bibfnamefont {P.}~\bibnamefont
  {Cui}}, \bibinfo {author} {\bibfnamefont {S.-M.}\ \bibnamefont {Li}},
  \bibinfo {author} {\bibfnamefont {S.}~\bibnamefont {Li}}, \bibinfo {author}
  {\bibfnamefont {Y.-L.}\ \bibnamefont {Liu}}, \ and\ \bibinfo {author}
  {\bibfnamefont {A.~M.}\ \bibnamefont {Zhang}},\ }\bibfield  {title} {\enquote
  {\bibinfo {title} {Vertically neutral collapse of a pulsating bubble at the
  corner of a free surface and a rigid wall},}\ }\href {\doibase
  10.1017/jfm.2023.292} {\bibfield  {journal} {\bibinfo  {journal} {Journal of
  Fluid Mechanics}\ }\textbf {\bibinfo {volume} {962}},\ \bibinfo {pages} {A28}
  (\bibinfo {year} {2023})}\BibitemShut {NoStop}%
\bibitem [{\citenamefont {Li}\ \emph {et~al.}(2019)\citenamefont {Li},
  \citenamefont {Zhang}, \citenamefont {Wang},\ and\ \citenamefont
  {Zhang}}]{li19}%
  \BibitemOpen
  \bibfield  {author} {\bibinfo {author} {\bibfnamefont {S.~M.}\ \bibnamefont
  {Li}}, \bibinfo {author} {\bibfnamefont {A.~M.}\ \bibnamefont {Zhang}},
  \bibinfo {author} {\bibfnamefont {Q.~X.}\ \bibnamefont {Wang}}, \ and\
  \bibinfo {author} {\bibfnamefont {S.}~\bibnamefont {Zhang}},\ }\bibfield
  {title} {\enquote {\bibinfo {title} {The jet characteristics of bubbles near
  mixed boundaries},}\ }\href {\doibase 10.1063/1.5112049} {\bibfield
  {journal} {\bibinfo  {journal} {Physics of Fluids}\ }\textbf {\bibinfo
  {volume} {31}},\ \bibinfo {pages} {107105} (\bibinfo {year}
  {2019})}\BibitemShut {NoStop}%
\bibitem [{\citenamefont {Vogel}, \citenamefont {Busch},\ and\ \citenamefont
  {Parlitz}(1996)}]{vogel96}%
  \BibitemOpen
  \bibfield  {author} {\bibinfo {author} {\bibfnamefont {A.}~\bibnamefont
  {Vogel}}, \bibinfo {author} {\bibfnamefont {S.}~\bibnamefont {Busch}}, \ and\
  \bibinfo {author} {\bibfnamefont {U.}~\bibnamefont {Parlitz}},\ }\bibfield
  {title} {\enquote {\bibinfo {title} {Shock wave emission and cavitation
  bubble generation by picosecond and nanosecond optical breakdown in water},}\
  }\href {\doibase 10.1121/1.415878} {\bibfield  {journal} {\bibinfo  {journal}
  {The Journal of the Acoustical Society of America}\ }\textbf {\bibinfo
  {volume} {100}},\ \bibinfo {pages} {148--165} (\bibinfo {year}
  {1996})}\BibitemShut {NoStop}%
\bibitem [{\citenamefont {Han}\ \emph {et~al.}(2015)\citenamefont {Han},
  \citenamefont {Jungnickel}, \citenamefont {K\"{o}hler}, \citenamefont
  {Lauterborn}, \citenamefont {Mettin},\ and\ \citenamefont {Vogel}}]{han15}%
  \BibitemOpen
  \bibfield  {author} {\bibinfo {author} {\bibfnamefont {B.}~\bibnamefont
  {Han}}, \bibinfo {author} {\bibfnamefont {K.}~\bibnamefont {Jungnickel}},
  \bibinfo {author} {\bibfnamefont {K.}~\bibnamefont {K\"{o}hler}}, \bibinfo
  {author} {\bibfnamefont {W.}~\bibnamefont {Lauterborn}}, \bibinfo {author}
  {\bibfnamefont {R.}~\bibnamefont {Mettin}}, \ and\ \bibinfo {author}
  {\bibfnamefont {A.}~\bibnamefont {Vogel}},\ }\bibfield  {title} {\enquote
  {\bibinfo {title} {Dynamics of laser-induced bubble pairs},}\ }\href
  {\doibase 10.1017/jfm.2015.183} {\bibfield  {journal} {\bibinfo  {journal}
  {Journal of Fluid Mechanics}\ }\textbf {\bibinfo {volume} {771}},\ \bibinfo
  {pages} {706--742} (\bibinfo {year} {2015})}\BibitemShut {NoStop}%
\bibitem [{\citenamefont {Lauterborn}\ and\ \citenamefont
  {Bolle}(1975)}]{lauterborn75}%
  \BibitemOpen
  \bibfield  {author} {\bibinfo {author} {\bibfnamefont {W.}~\bibnamefont
  {Lauterborn}}\ and\ \bibinfo {author} {\bibfnamefont {H.}~\bibnamefont
  {Bolle}},\ }\bibfield  {title} {\enquote {\bibinfo {title} {Experimental
  investigations of cavitation-bubble collapse in the neighbourhood of a solid
  boundary},}\ }\href {\doibase 10.1017/S0022112075003448} {\bibfield
  {journal} {\bibinfo  {journal} {Journal of Fluid Mechanics}\ }\textbf
  {\bibinfo {volume} {72}},\ \bibinfo {pages} {391--399} (\bibinfo {year}
  {1975})}\BibitemShut {NoStop}%
\bibitem [{\citenamefont {Plesset}\ and\ \citenamefont
  {Chapman}(1971)}]{plesset71}%
  \BibitemOpen
  \bibfield  {author} {\bibinfo {author} {\bibfnamefont {M.~S.}\ \bibnamefont
  {Plesset}}\ and\ \bibinfo {author} {\bibfnamefont {R.~B.}\ \bibnamefont
  {Chapman}},\ }\bibfield  {title} {\enquote {\bibinfo {title} {Collapse of an
  initially spherical vapour cavity in the neighbourhood of a solid
  boundary},}\ }\href {\doibase 10.1017/S0022112071001058} {\bibfield
  {journal} {\bibinfo  {journal} {Journal of Fluid Mechanics}\ }\textbf
  {\bibinfo {volume} {47}},\ \bibinfo {pages} {283--290} (\bibinfo {year}
  {1971})}\BibitemShut {NoStop}%
\bibitem [{\citenamefont {Reuter}\ and\ \citenamefont
  {Kaiser}(2019)}]{reuter19}%
  \BibitemOpen
  \bibfield  {author} {\bibinfo {author} {\bibfnamefont {F.}~\bibnamefont
  {Reuter}}\ and\ \bibinfo {author} {\bibfnamefont {S.~A.}\ \bibnamefont
  {Kaiser}},\ }\bibfield  {title} {\enquote {\bibinfo {title} {High-speed
  film-thickness measurements between a collapsing cavitation bubble and a
  solid surface with total internal reflection shadowmetry},}\ }\href {\doibase
  10.1063/1.5095148} {\bibfield  {journal} {\bibinfo  {journal} {Physics of
  Fluids}\ }\textbf {\bibinfo {volume} {31}},\ \bibinfo {pages} {097108}
  (\bibinfo {year} {2019})}\BibitemShut {NoStop}%
\bibitem [{\citenamefont {Zeng}, \citenamefont {Gonzalez-Avila},\ and\
  \citenamefont {Ohl}(2020)}]{zeng20}%
  \BibitemOpen
  \bibfield  {author} {\bibinfo {author} {\bibfnamefont {Q.}~\bibnamefont
  {Zeng}}, \bibinfo {author} {\bibfnamefont {S.~R.}\ \bibnamefont
  {Gonzalez-Avila}}, \ and\ \bibinfo {author} {\bibfnamefont {C.-D.}\
  \bibnamefont {Ohl}},\ }\bibfield  {title} {\enquote {\bibinfo {title}
  {Splitting and jetting of cavitation bubbles in thin gaps},}\ }\href
  {\doibase 10.1017/jfm.2020.356} {\bibfield  {journal} {\bibinfo  {journal}
  {Journal of Fluid Mechanics}\ }\textbf {\bibinfo {volume} {896}},\ \bibinfo
  {pages} {A28} (\bibinfo {year} {2020})}\BibitemShut {NoStop}%
\bibitem [{\citenamefont {Vogel}, \citenamefont {Lauterborn},\ and\
  \citenamefont {Timm}(1989)}]{vogel89}%
  \BibitemOpen
  \bibfield  {author} {\bibinfo {author} {\bibfnamefont {A.}~\bibnamefont
  {Vogel}}, \bibinfo {author} {\bibfnamefont {W.}~\bibnamefont {Lauterborn}}, \
  and\ \bibinfo {author} {\bibfnamefont {R.}~\bibnamefont {Timm}},\ }\bibfield
  {title} {\enquote {\bibinfo {title} {Optical and acoustic investigations of
  the dynamics of laser-produced cavitation bubbles near a solid boundary},}\
  }\href {\doibase 10.1017/S0022112089002314} {\bibfield  {journal} {\bibinfo
  {journal} {Journal of Fluid Mechanics}\ }\textbf {\bibinfo {volume} {206}},\
  \bibinfo {pages} {299--338} (\bibinfo {year} {1989})}\BibitemShut {NoStop}%
\bibitem [{\citenamefont {Blake}(1988)}]{blake1988kelvin}%
  \BibitemOpen
  \bibfield  {author} {\bibinfo {author} {\bibfnamefont {J.~R.}\ \bibnamefont
  {Blake}},\ }\bibfield  {title} {\enquote {\bibinfo {title} {The kelvin
  impulse: application to cavitation bubble dynamics},}\ }\href@noop {}
  {\bibfield  {journal} {\bibinfo  {journal} {The ANZIAM Journal}\ }\textbf
  {\bibinfo {volume} {30}},\ \bibinfo {pages} {127--146} (\bibinfo {year}
  {1988})}\BibitemShut {NoStop}%
\bibitem [{\citenamefont {Obreschkow}\ \emph {et~al.}(2011)\citenamefont
  {Obreschkow}, \citenamefont {Tinguely}, \citenamefont {Dorsaz}, \citenamefont
  {Kobel}, \citenamefont {de~Bosset},\ and\ \citenamefont
  {Farhat}}]{Obreschkow2011gravity}%
  \BibitemOpen
  \bibfield  {author} {\bibinfo {author} {\bibfnamefont {D.}~\bibnamefont
  {Obreschkow}}, \bibinfo {author} {\bibfnamefont {M.}~\bibnamefont
  {Tinguely}}, \bibinfo {author} {\bibfnamefont {N.}~\bibnamefont {Dorsaz}},
  \bibinfo {author} {\bibfnamefont {P.}~\bibnamefont {Kobel}}, \bibinfo
  {author} {\bibfnamefont {A.}~\bibnamefont {de~Bosset}}, \ and\ \bibinfo
  {author} {\bibfnamefont {M.}~\bibnamefont {Farhat}},\ }\bibfield  {title}
  {\enquote {\bibinfo {title} {Universal scaling law for jets of collapsing
  bubbles},}\ }\href {\doibase 10.1103/PhysRevLett.107.204501} {\bibfield
  {journal} {\bibinfo  {journal} {Physical Review Letters}\ }\textbf {\bibinfo
  {volume} {107}},\ \bibinfo {pages} {204501} (\bibinfo {year}
  {2011})}\BibitemShut {NoStop}%
\bibitem [{\citenamefont {Brujan}\ \emph {et~al.}(2022)\citenamefont {Brujan},
  \citenamefont {Zhang}, \citenamefont {Liu}, \citenamefont {Ogasawara},\ and\
  \citenamefont {Takahira}}]{brujan22}%
  \BibitemOpen
  \bibfield  {author} {\bibinfo {author} {\bibfnamefont {E.-A.}\ \bibnamefont
  {Brujan}}, \bibinfo {author} {\bibfnamefont {A.~M.}\ \bibnamefont {Zhang}},
  \bibinfo {author} {\bibfnamefont {Y.-L.}\ \bibnamefont {Liu}}, \bibinfo
  {author} {\bibfnamefont {T.}~\bibnamefont {Ogasawara}}, \ and\ \bibinfo
  {author} {\bibfnamefont {H.}~\bibnamefont {Takahira}},\ }\bibfield  {title}
  {\enquote {\bibinfo {title} {Jetting and migration of a laser-induced
  cavitation bubble in a rectangular channel},}\ }\href {\doibase
  10.1017/jfm.2022.695} {\bibfield  {journal} {\bibinfo  {journal} {Journal of
  Fluid Mechanics}\ }\textbf {\bibinfo {volume} {948}},\ \bibinfo {pages} {A6}
  (\bibinfo {year} {2022})}\BibitemShut {NoStop}%
\end{thebibliography}%
\end{document}